\documentclass[a4paper,fleqn,usenatbib]{mnras}

\usepackage{pdflscape}
\usepackage{newtxtext,newtxmath}
\usepackage[T1]{fontenc}
\usepackage{ae,aecompl}
\usepackage{graphicx}	% Including figure files
\usepackage{amsmath}	% Advanced maths commands
\usepackage{amssymb}	% Extra maths symbols
\defcitealias{arenou}{AGG}
\defcitealias{sfd}{SFD}
\defcitealias{drimmel}{DCL}
\defcitealias{av}{G12}
\defcitealias{g17}{G17}
\defcitealias{lallement2014}{LVV}

\title[Polarisation and extinction]
{Interstellar polarisation and extinction in the Local Bubble and the Gould Belt}

\author[G. A. Gontcharov and A. V. Mosenkov]{
George A. Gontcharov,$^{1,2}$\thanks{E-mail: george.gontcharov@tdt.edu.vn}
and Aleksandr V. Mosenkov,$^{3}$
\\
$^{1}$Department for Management of Science and Technology Development, Ton Duc Thang University, Ho Chi Minh City, Vietnam\\
$^{2}$Faculty of Applied Sciences, Ton Duc Thang University, Ho Chi Minh City, Vietnam\\
$^{3}$Central Astronomical Observatory, Russian Academy of Sciences, 65/1 Pulkovskoye chaussee, St. Petersburg, 196140 Russia
}

\date{Accepted 2018 October 29. Received 2018 October 25; in original form 2018 February 16}

\pubyear{2018}

\begin{document}
\label{firstpage}
\pagerange{\pageref{firstpage}--\pageref{lastpage}}
\maketitle

\begin{abstract}
We analyse an all-sky compilation of thirteen data sources with optical interstellar linear polarisation for 
3871 {\it Gaia} DR2 and {\it Hipparcos} stars
within 500 pc, which do not exhibit a considerable intrinsic polarisation.
These data are analysed together with five 3D maps and models of the reddening $E(B-V)$.
We consider variations
of the polarisation degree $P$, position angle $\theta$, and 
polarisation efficiency $P/E(B-V)$ 
with Galactic coordinates, distance $R$, and dereddened colour.
$P$ shows a maximum at the Gould Belt mid-plane.
$P/R$ drops in the Local Bubble by several times defining a boundary of the Bubble at $P=0.1$ per cent.
All the data sources of the reddening, except Lallement et al., show a drop of $P/E(B-V)$ in the Bubble.
A significant number of stars with a too high $P/E(B-V)$, where the reddening is taken from Arenou et al.,
Drimmel et al. or Lallement et al., either reject the Serkowski's limit $P/E(B-V)<9.3$ per cent per mag
or manifest a systematic underestimation of the reddening by these data sources.
The spatial and colour dependent variations of $P$ and $E(B-V)$ outside the Bubble compensate each other,
resulting in a nearly constant $P/E(B-V)$.
A giant envelope of aligned dust dominates at middle and high latitudes outside the Bubble.
The Markkanen's cloud, the North Polar Spur, and some other filaments are parts of this envelope.
\end{abstract}


\section{Introduction}
\label{intro}

Cumulative interstellar extinction to a star and its reddening are measures of the optical depth of the 
interstellar dust medium.
On the other hand, interstellar polarisation of a star is a measure of the alignment of the dust grains by a 
strong and regular magnetic field between the observer and the star.
Thus, interstellar polarisation, combined with interstellar extinction and reddening, as well as with 
multi-colour photometry and high-precision distances, offers a possibility to study the local distribution 
and properties of interstellar dust and Galactic magnetic field.

The position angle of polarisation vectors can be a particularly appropriate tool to map the local Galactic 
magnetic field \citep{berd2014}.
It is generally accepted that this field is directed mainly along the Local spiral arm, i.e. nearly
along $l\approx(80\degr, 260\degr)$ \citep{stephens2011}. 
However, local deviations of the field are known \citep{leroy1999} and should be studied in detail.

Interstellar extinction and interstellar polarisation must be separated from 
intrinsic extinction of a star (usually negligible) and its intrinsic polarisation (sometimes dominating).
It is a common practice to determine a value for the interstellar polarisation of a star by taking 
into account an average or median of the measured polarisation of nearby stars,
which do not exhibit a considerable intrinsic polarisation
\citep{cotton2017}. 
Therefore, some detailed 3D maps of the spatial variations of the polarisation degree and position angle are 
highly needed.
Up to now, there have been made only some initial attempts of creating such maps
\citep{fosalba2002, stephens2011, santos2011} .

Polarisation has been measured for thousands of stars and presented in a dozen of catalogues \citep{heiles2000}. 
Some previous attempts to combine these data \citep{leroy1999, heiles2000, fosalba2002} have shown that the Local Bubble
(or Local Void, or Local Cavity) within nearly 100 pc from the Sun
is a space with virtually no polarisation,  whereas it is considerable outside the Bubble.
The Bubble is known as a region of lower interstellar gas density \citep{welsh2010} surrounded by a torus of
denser gas known, in its turn, as the Gould Belt.
It is believed that the Bubble was formed few Myr ago by one or more supernovae 
and since then there has been no star formation in it \citep{abt2011}.
In contrast, the Belt is a region of ongoing star formation \citep{bobylev2014}.
The Bubble is opened in the direction of the Galactic halo, forming a `chimney' nearly
perpendicular to the plane of the Belt \citep{welsh1999}.
Being tilted to the Galactic mid-plane at about 20$\degr$, the Belt takes maximal and minimal Galactic latitudes 
near the directions of the Galactic Centre and anticentre, respectively.
The Belt extends up to several hundred parsecs from the Sun.

Despite the agreement on the general structure of the nearby space 
\citep[see][ pp. 324--328]{reis2011, astroph, perr},
the quantitative characteristics of the medium in the Bubble and the Belt remain contradictory.
Beginning with \citet{fitzgerald1968}, it is believed that the Bubble has lower dust density, reddening, and extinction
in addition to lower gas density \citep[see][]{reis2008, bobylev2014, lallement2014, frisch2015, capitanio2017}.
However, in fact, \citeauthor{fitzgerald1968} only drew a nearby region with a total reddening $E(B-V)<0.01$~mag.
An increase of the reddening outside the Bubble may be a simple consequence of the natural increase of the dust column 
density with distance when its volume density is nearly constant.
\citeauthor{fitzgerald1968} did not show that this volume density and related differential reddening $E(B-V)/R$ is 
lower near the Sun ($R$ is the distance).
This can only be tested with fairly accurately estimated distances in the post-{\it Hipparcos} era.
For example, figure~1 from \citet{lallement2014} and figure~4 from \citet{capitanio2017} 
do show the Bubble as a region with lower dust volume density.
However, 
various definitions of the Bubble by estimating $E(B-V)$ at its boundary are contradictory:
$E(B-V)<0.010$ \citep{fitzgerald1968},
$E(B-V)<0.045$ \citep{frisch2007},
$E(B-V)<0.053$ given $E(b-y)<0.04$ and $E(B-V)=1.335\,E(b-y)$ \citep{lallement2014}, and even
$E(B-V)<0.055$ given the extinction-to-reddening ratio $R_\mathrm{V}\equiv A_\mathrm{V}/E(B-V)=3.1$ and 
$A_\mathrm{V}<0.17$~mag \citep{reis2011}.
A typical accuracy of all these estimates seems to be $\sigma(E(B-V))>0.02$~mag \citep{gm2017big}.
With such a high uncertainty, it is still to be confirmed, whether the Bubble has lower dust volume density, 
$E(B-V)/R$ and $A_\mathrm{V}/R$.

The low polarisation, reddening, and extinction inside the Bubble, 
together with their low fractional accuracy, have led to some contradictory statements.
For example, \citet{cotton2016} noted that `at 100 pc the extinction is well below that which can actually 
be measured by photometry', in contradiction to a note by \citet{leroy1999} that 
`The walls of the Bubble [closer than 100 pc] are fairly well defined when one measures the interstellar absorption'.
A comparison of polarisation and reddening for the same stars may solve such contradictions.

Some attempts to give a mean relationship between the polarisation degree and the reddening have been 
based on the data with either a poor statistics or low accuracy. 
For example, \citet{fosalba2002} provided the mean relationship
\begin{equation}
\label{fosa}
P=3.5\,E(B-V)^{0.8}~per~cent\,,
\end{equation}
based on an inhomogeneous compilation of $E(B-V)$, which are precise to only $\sigma(E(B-V))=0.1$~mag.
Moreover, due to a rather sparse distribution of the stars, \citeauthor{fosalba2002} had to derive a single equation 
for the whole space of several kiloparsecs.
A very large scatter of these data (figure~4 from \citeauthor{fosalba2002}) suggests 
that there is a large variety of such relations in different regions of space.
Some of our results are compared with those from \citeauthor{fosalba2002} in Section~\ref{colorbl}.

In previous comparisons of polarisation and reddening since \citet{serk1975}, 
the maximum value of the polarisation efficiency 
$P/A_\mathrm{V}\approx3$, i.e. $P/E(B-V)\approx9.3$ per cent per mag with $R_\mathrm{V}=3.1$
has been found empirically (hereafter Serkowski's limit).
However, it is based on a rather poor statistics: for example, \citeauthor{serk1975} used only less 
than 300 stars, whereas in recent study \citet{planck21} used only 206 objects.
Stars with $P/E(B-V)>9.3$ per cent per mag tend to be considered as suspected intrinsically polarised stars.
However, for most of them no indication of an intrinsic polarisation has been found.
It is thus worthwhile to verify the Serkowski's limit and $P/E(B-V)$ variations
by use of most reliable all-sky reddening data sources.

Today we have data to make more robust conclusions about the spatial variations and relations of interstellar 
polarisation and reddening/extinction inside and outside the Bubble.
These data include new polarisation measurements, reddening data from new sources based on some
recent large all-sky photometric and spectroscopic surveys, as well 
as more accurate parallaxes 
and photometry from the {\it Gaia} DR2 \citep{gaiabrown, gaiaevans, gaialuri}.
Note that \citet{heiles2000} and \citet{fosalba2002} are the latest all-sky compilation of the polarisation data 
and all-sky analysis of their spatial variations, respectively.
Thus, the potential of very accurate {\it Gaia} parallaxes has not been used.

A new dataset and method of \citet[][hereafter LVV]{lallement2014} are especially promising for the nearby 
space with low reddening.
`One of the main advantages of the present method and of the dataset is the large number of nearby stars for 
which very low extinctions have been measured' \citepalias{lallement2014}.

The aim of our study is to compile an all-sky polarisation dataset in- and outside the Bubble, to 
show this dataset versus the most precise reddening estimates and to study some variations of 
$P$, $\theta$, $E(B-V)$, and $P/E(B-V)$ as some functions of the Galactic coordinates $R$, $l$, and $b$, and stellar colour.

Besides some global variations, we intend to test some minor assumptions.
For example, a higher polarisation has been found for stars of the spectral classes B and M \citep[see][]{cotton2016}.
However,
in all the samples under consideration these classes are more luminous and distant.
Therefore, the higher polarisation 
of these classes may be only a consequence of the polarisation versus distance relation \citep{bailey2010}.

This paper is organised as follows. In the next section, we present polarimetric data and analyse
spatial variations of the polarisation degree and position angle.
In Section~\ref{versus}, we analyse the polarisation degree together with the reddening and 
polarisation efficiency as some functions of Galactic coordinates and dereddened colour of stars.
Our conclusions are presented in Section~\ref{conclusions}.

\begin{figure*}
\includegraphics{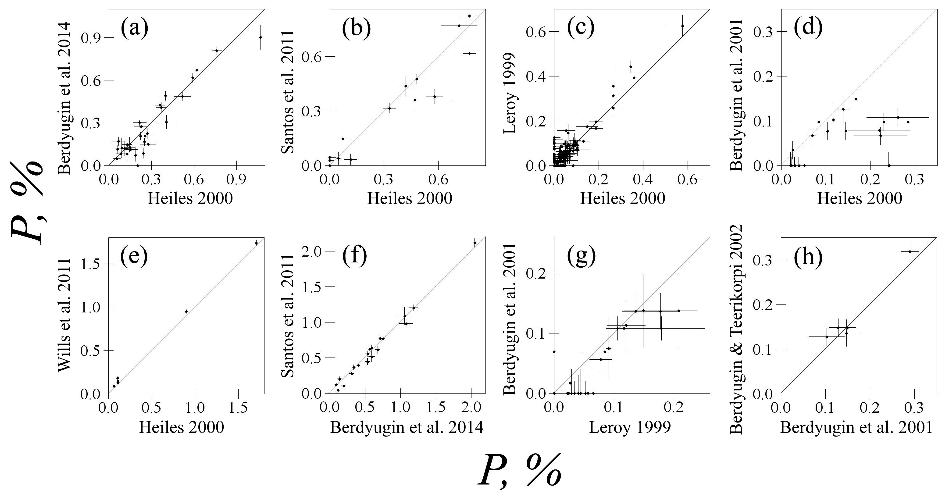}
\caption{Polarisation degree (per cent) for common stars in the pairs of the data sources:
(a) \citet{heiles2000} versus \citet{berd2014},
(b) \citet{heiles2000} versus \citet{santos2011},
(c) \citet{heiles2000} versus \citet{leroy1999},
(d) \citet{heiles2000} versus \citet{berd2001},
(e) \citet{heiles2000} versus \citet{wills2011},
(f) \citet{berd2014} versus \citet{santos2011},
(g) \citet{leroy1999} versus \citet{berd2001},
(h) \citet{berd2001} versus \citet{berd2002}.
}
\label{pp}
\end{figure*}

\begin{figure*}
\includegraphics{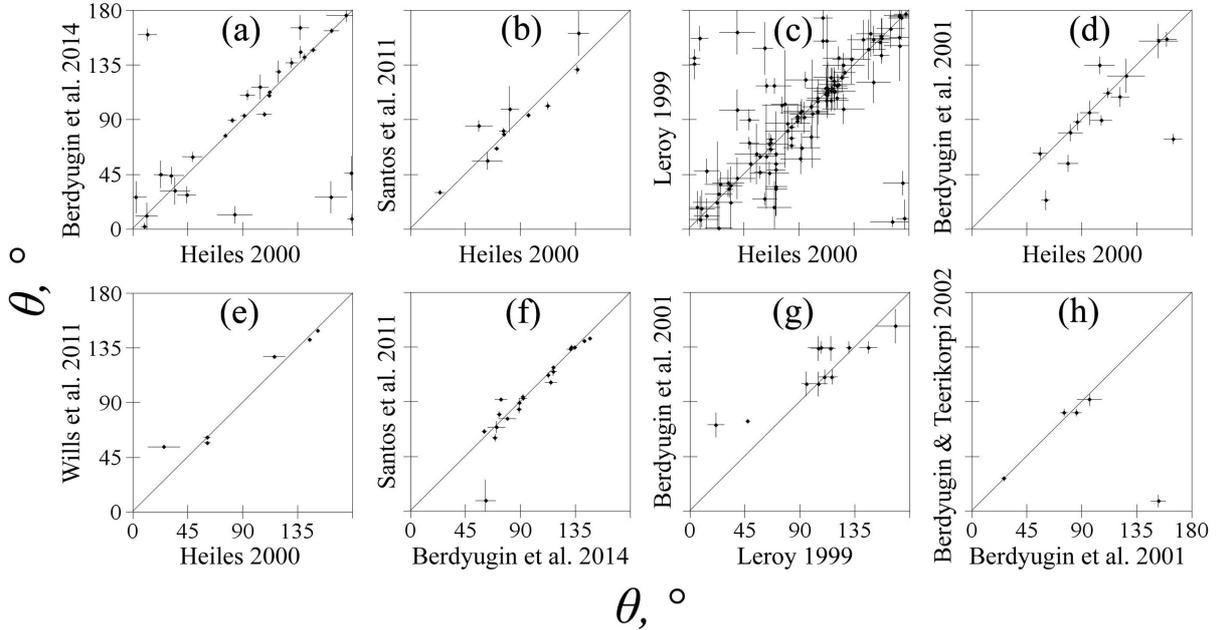}
\caption{The same as Fig.~\ref{pp} but for the polarisation position angle (in degrees) for common stars.
}
\label{papa}
\end{figure*}

\section{Polarisation}
\label{polarisation}

Throughout this work, the polarisation degree $P$ and position angle $\theta$ are given in per cents and degrees, 
respectively.
All measurements of $\theta$ are referred to Galactic coordinates \citep{stephens2011}.

In order to analyse spatial variations of polarisation and reddening, we have to consider stars with precise parallaxes.
This is especially important because all reddening data sources used give reddening as some functions of 
3D coordinates.
We select {\it Gaia} DR2 stars with precise measurements of optical linear polarisation and 
within 500 pc based on the distances from \citet{bailer2018}.
Also, we select 55 bright stars with precise polarisation measurements, which have precise {\it Hipparcos}
parallaxes \citep{hip2} but not included in {\it Gaia} DR2.
The median fractional parallax uncertainty of the sample is very low: $\sigma(\varpi)/\varpi=0.01$ and, hence,
has a negligible contribution to other uncertainties.
21 stars with $\sigma(\varpi)/\varpi>0.2$ are eliminated due to their very uncertain $R$ \citep{bailer2015}.
The space under consideration includes the Local Bubble and nearly the whole Gould Belt.
Yet, a rather sparse distribution of stars with a measured polarisation at the edge of this space does not allow 
us to detect the outer boundary of the Belt by any characteristic.

We consider optical polarisation measurements from 13 data sources:
\citet{heiles2000} -- 1595 independent values in the final sample,
\citet{berd2014} -- 1290,
\citet{santos2011} -- 451,
\citet{leroy1999} -- 350,
\citet{berd2001} -- 97,
\citet{berd2002} -- 76,
\citet{alves2006} -- 62,
\citet{cotton2017} -- 45,
\citet{wills2011} -- 31,
\citet{bailey2010} -- 23,
\citet{cotton2016} -- 21,
\citet{poidevin2006} -- 20,
and \citet{andersson2007} -- 6 values
after the elimination of some  values/stars from the sample, as described below.
Some stars have several independent values.
Therefore, the number of stars under consideration is slightly lower than the number of values.

To consider the polarisation statistics, we de-biased the polarisation degree estimates, following a standard 
practice \citep[see][]{cotton2017},
as:
\begin{equation}
\label{debias}
P_{debiased}=(P^2-\sigma(P)^2)^{1/2}\,,
\end{equation}
as $P>\sigma(P)$ for our sample.

%¹¹¹¹¹¹¹¹¹¹¹¹¹¹¹¹¹¹¹¹¹¹¹¹¹¹¹¹¹
\begin{table*}
\def\baselinestretch{1}\normalsize\scriptsize
\caption[]{The master table with polarisation and reddening data:
the {\it Gaia} DR2 source\_id or other name, {\it Hipparcos} number for stars not found in {\it Gaia} DR2, polarisation
data source, distance (parsecs), Galactic coordinates (degrees),
$P$ and its $\sigma$ (per cents), $\theta$ and its $\sigma$ (degrees),
$R_\mathrm{V}$, and $E(B-V)$ from \citetalias{g17}, \citetalias{av}, \citetalias{arenou}, \citetalias{drimmel}, and
\citetalias{lallement2014}. The complete table is available online.
}
\label{master}
\[
\begin{tabular}{rrlrrrrrrrrrrrrr}
\hline
\noalign{\smallskip}
DR2 Source\_id / Name & HIP & Data source & $R$ & $l$ & $b$ & $P$ & $\sigma(P)$ & $\theta$ & $\sigma(\theta)$ & $R_\mathrm{V}$ & G17 & G12 & AGG & DCL & LVV  \\
\hline
\noalign{\smallskip}
1002767184492235648 & & Heiles2000 & 108.8 & 156.3675 & 25.5549 & 0.40000 & 0.10000 & 176.7 &  7.2 & 3.10 & 0.066 & 0.061 & 0.016 & 0.011 & 0.004 \\
1002792679418093696 & & Heiles2000 & 280.2 & 156.2458 & 25.7459 & 0.60000 & 0.10000 & 176.4 &  4.8 & 2.93 & 0.068 & 0.098 & 0.033 & 0.049 & 0.049 \\
1008921872627395200 & & Berd2014   & 336.5 & 178.3186 & 41.1850 & 0.08100 & 0.02400 & 102.7 &  8.5 & 2.77 & 0.059 & 0.084 & 0.018 & 0.020 & 0.012 \\
1009508393361454464 & & Berd2014   & 217.6 & 175.6596 & 42.1903 & 0.03500 & 0.01800 &  72.1 & 14.7 & 3.14 & 0.041 & 0.069 & 0.018 & 0.014 & 0.007 \\
1009982970068091904 & & Berd2014   & 286.1 & 176.5244 & 40.4675 & 0.08300 & 0.02400 & 106.7 &  8.3 & 2.80 & 0.039 & 0.082 & 0.018 & 0.020 & 0.011 \\
\ldots & \ldots & \ldots & \ldots & \ldots & \ldots & \ldots & \ldots & \ldots & \ldots & \ldots & \ldots & \ldots & \ldots & \ldots & \ldots \\
\hline
\end{tabular}
\]
\end{table*}

%¹¹¹¹¹¹¹¹¹¹¹¹¹¹¹¹¹¹¹¹¹¹¹¹¹¹¹¹¹¹¹¹¹¹¹¹

\subsection{Verification of the sample}
\label{verification}

In order to clean and verify the polarisation data, we use all the data with $P/\sigma(P)>1.43$.
These data together with the reddenings are presented in the master Table~~\ref{master} available on-line.
After this verification we select the final sample with $P/\sigma(P)>3$ in order to make more robust conclusions.
This restriction translates into $\sigma(\theta)<9.55\degr$ following \citet{serk1974}.

The polarisation measurements were fulfilled in the $B$, $V$, SDSS $g'$, a wide visual band,
or have been scaled to these bands by the authors of these publications.
No author found significant variations of $P$ or $\theta$ within these 
bands \citep{berd2001, alves2006, andersson2007, berd2014, cotton2016}.
Thus, we can combine these data into one sample.

In our sample we excluded stars with intrinsic polarisation known from the literature.
These include all Be stars from \citet{yudin2001}, 
all stars with IR excess from \citet{mcdonald2017}, except {\it Gaia} DR2 3520586071216143488
and {\it Gaia} DR2 4076915349748285952 with negligible intrinsic polarisation \citep{cotton2017},
as well as several stars from other sources.

%¹¹¹¹¹¹¹¹¹¹¹¹¹¹¹¹¹¹¹¹¹¹¹¹¹¹¹¹¹
\begin{table*}
\def\baselinestretch{1}\normalsize\normalsize
\caption[]{The 135 excluded stars with a suspected intrinsic polarisation:
the {\it Gaia} DR2 source\_id, $G$ magnitude, distance (parsecs), Galactic coordinates (degrees), spectrum, and notes.
The complete table is available online.
}
\label{suspectintrinsic}
\[
\begin{tabular}{lrrrrll}
\hline
\noalign{\smallskip}
{\it Gaia} DR2 Source\_id & $G$ & $R$ & $l$ & $b$ & Spectrum & Note \\
\hline
\noalign{\smallskip}
1053778957742409984 & 4.90 & 101 & 143.5408 & +45.2202 & A1:VpSiSrHg  & ET~UMa, Variable of alpha2 CVn type \\
1084415852819290112 & 5.99 & 89  & 156.7298 & +31.9494 & A3VpSrSiCrEu & 53~Cam, Variable of alpha2 CVn type, Binary \\
1166857200309419392 & 7.53 & 181 &  10.2360 & +52.0205 & A3p          & LV~Ser, Variable of alpha2 CVn type \\
1190927159109415552 & 5.29 & 69  &  21.8345 & +47.7510 & A2VpMnEu(Sr) & chi~Ser, Variable of alpha2 CVn type, Disagree \\
1197801408889764480 & 5.76 & 54  &  29.6874 & +49.9625 & A8Vam        & 22~Ser, Binary \\
\ldots & \ldots & \ldots & \ldots & \ldots & \ldots & \ldots \\
\hline
\end{tabular}
\]
\end{table*}

%¹¹¹¹¹¹¹¹¹¹¹¹¹¹¹¹¹¹¹¹¹¹¹¹¹¹¹¹¹¹¹¹¹¹¹¹

Also, we exclude 135 stars with a suspected intrinsic polarisation. They are listed in Table~\ref{suspectintrinsic}.
The reasons for excluding these stars are the following: some specific features in their spectral classification
(e.g. Ap), duplicity, variability, fast rotation, or high proper motion (probably, followed by misidentification).
This information is derived from the SIMBAD database \citep{simbad}, the 
Tycho-2 Spectral Type Catalog \citep{tst},
the Washington Visual Double Star Catalog \citep{wds},
and other sources.
Few stars are excluded due to their discrepant degree of polarisation relative to neighbouring stars.
These rejected stars are designated in the notes in the Table~\ref{suspectintrinsic} as having `Too high polarisation'.
Another reason is a disagreement of more than $3\sigma$ of the stated errors between the independent polarisation 
measurements for the same star.
Such cases are designated in the notes as `Disagree'.
Only 10 stars among them have such a disagreement as the main reason for a suspected intrinsic polarisation.

For the remaining stars with independent measurements (with a disagreement of less than $3\sigma$ of the stated errors)
Fig.~\ref{pp} and \ref{papa} show a comparison 
of $P$ and $\theta$, respectively, for the pairs of independent measurements.
We note that the limitation $0\degr<\theta<180\degr$, by definition, means that, in fact,
$\theta\approx0\degr$ agrees with $\theta\approx180\degr$.
A good agreement, which is evident from Fig.~\ref{pp} and \ref{papa}, allows us to compile all the measurements together.
We note that some earlier reported disagreement by $\theta$ in different data sources, for example by
\citet{alves2006}, is due to its ignored or improper conversion from equatorial to Galactic coordinates.
86 stars with an agreement within $1\sigma$ of the stated errors for $P$ and 
within 19$\degr$ for $\theta$, having $P/\sigma(P)>3$, are listed in Table~\ref{beststars}.

%¹¹¹¹¹¹¹¹¹¹¹¹¹¹¹¹¹¹¹¹¹¹¹¹¹¹¹¹¹
\begin{table*}
\def\baselinestretch{1}\normalsize\normalsize
\caption[]{The 86 stars with agreed independent polarisation measurements:
the {\it Gaia} DR2 source\_id, $G$ magnitude, distance (parsecs), Galactic coordinates (degrees), 
weighted mean $P$ and its $\sigma$ (per cents), weighted mean $\theta$ and its $\sigma$ (degrees).
The complete table is available online.
}
\label{beststars}
\[
\begin{tabular}{lrrrrrrrr}
\hline
\noalign{\smallskip}
{\it Gaia} DR2 Source\_id & $G$ & $R$ & $l$ & $b$ & Mean $P$ & $\sigma(P)$ & Mean $\theta$ & $\sigma(\theta)$ \\
\hline
\noalign{\smallskip}
1163440089969099264 & 8.61 & 218 &   9.0317 &  51.1356 & 0.265 & 0.011 & 137.4 & 1.2 \\
1250169120491310720 & 4.33 & 140 &  12.2481 &  75.4886 & 0.106 & 0.008 & 131.0 & 2.2 \\
1301081662819876480 & 8.03 & 338 &  44.8817 &  39.2163 & 0.268 & 0.019 & 163.2 & 2.1 \\
1501901345598817792 & 6.79 & 120 &  98.4798 &  70.7932 & 0.125 & 0.017 &  93.1 & 4.2 \\
2455241320597417856 & 4.40 &  84 & 157.0765 & -75.1550 & 0.019 & 0.005 &  93.4 & 6.9 \\
\ldots & \ldots & \ldots & \ldots & \ldots & \ldots & \ldots & \ldots & \ldots \\
\hline
\end{tabular}
\]
\end{table*}

%¹¹¹¹¹¹¹¹¹¹¹¹¹¹¹¹¹¹¹¹¹¹¹¹¹¹¹¹¹¹¹¹¹¹¹¹

\begin{figure}
\includegraphics{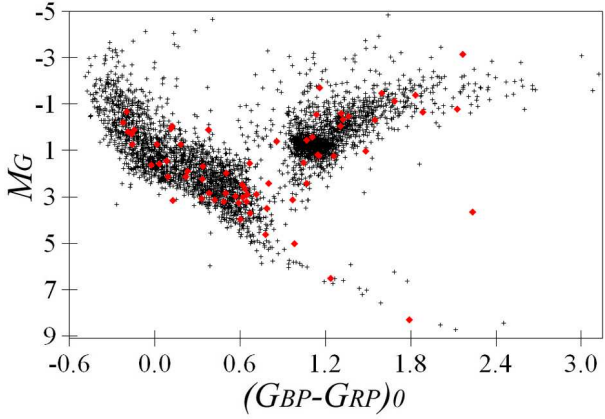}
\caption{The HR diagram for the final sample. 
The 67 stars with $P/E(B-V)>9.3$ per cent per mag from all the reddening data sources are shown with the red diamonds.
}
\label{hr}
\end{figure}

\begin{figure}
\includegraphics{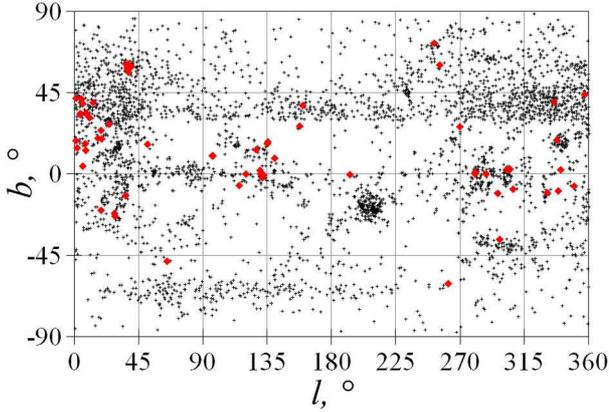}
\caption{The distribution of the final sample in the sky in Galactic coordinates. 
The 67 stars with $P/E(B-V)>9.3$ per cent per mag from all the reddening data sources are shown with the red diamonds.
}
\label{lb}
\end{figure}

It is likely that our sample still comprises some stars with a considerable but yet unknown intrinsic polarisation. 
Nevertheless, we believe that this would not alter the general pattern of the variations revealed in our study. 
We have tested all our results with and without the excluded stars -- their influence on the obtained results
is very small.

\begin{figure*}
\includegraphics{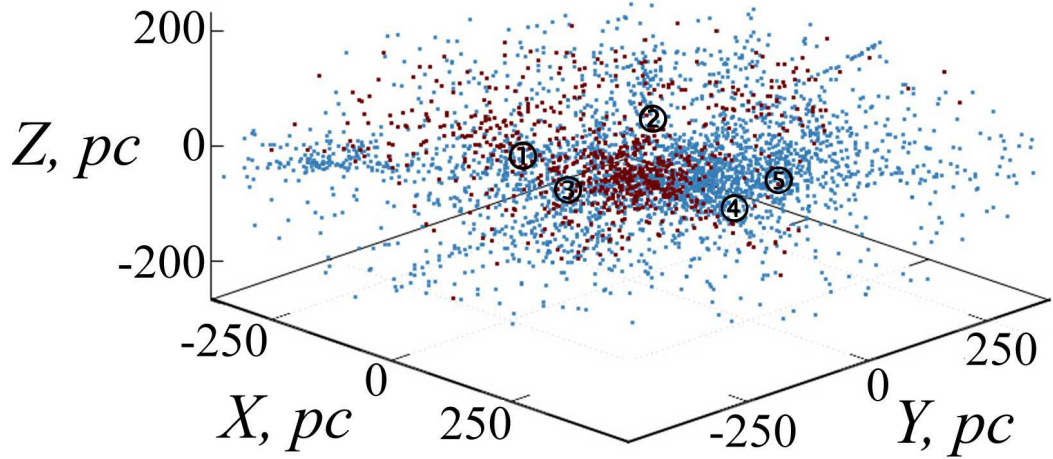}
\caption{The 3D distribution of the sample in the Galactic $XYZ$ coordinates.
The stars with $P<0.1$ and $>0.1$ per cent are shown with the brown and blue symbols.
The position of some bright stars is shown by the following digits:
1 -- Betelgeuse, 
2 -- Polaris, 
3 -- Canopus,
4 -- Antares, and
5 -- Enif.
}
\label{xyz}
\end{figure*}

\begin{figure}
\includegraphics{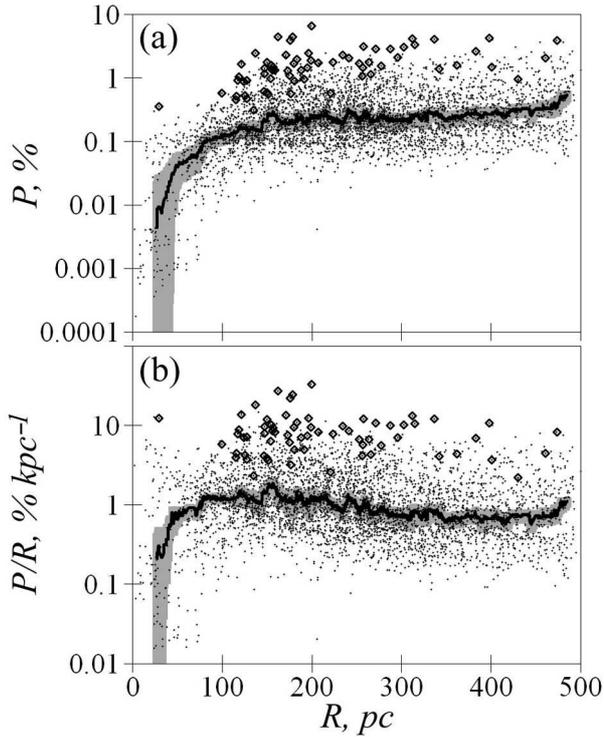}
\caption{$R$ (parsecs) versus (a) $P$ (per cent) and (b) $P/R$ (per cent per kiloparsec).
The 67 stars with $P/E(B-V)>9.3$ per cent per mag from all the reddening data sources are shown with the open diamonds.
The black curves with the grey error belts are the moving medians over 101 points.
}
\label{pol}
\end{figure}

\begin{figure}
\includegraphics{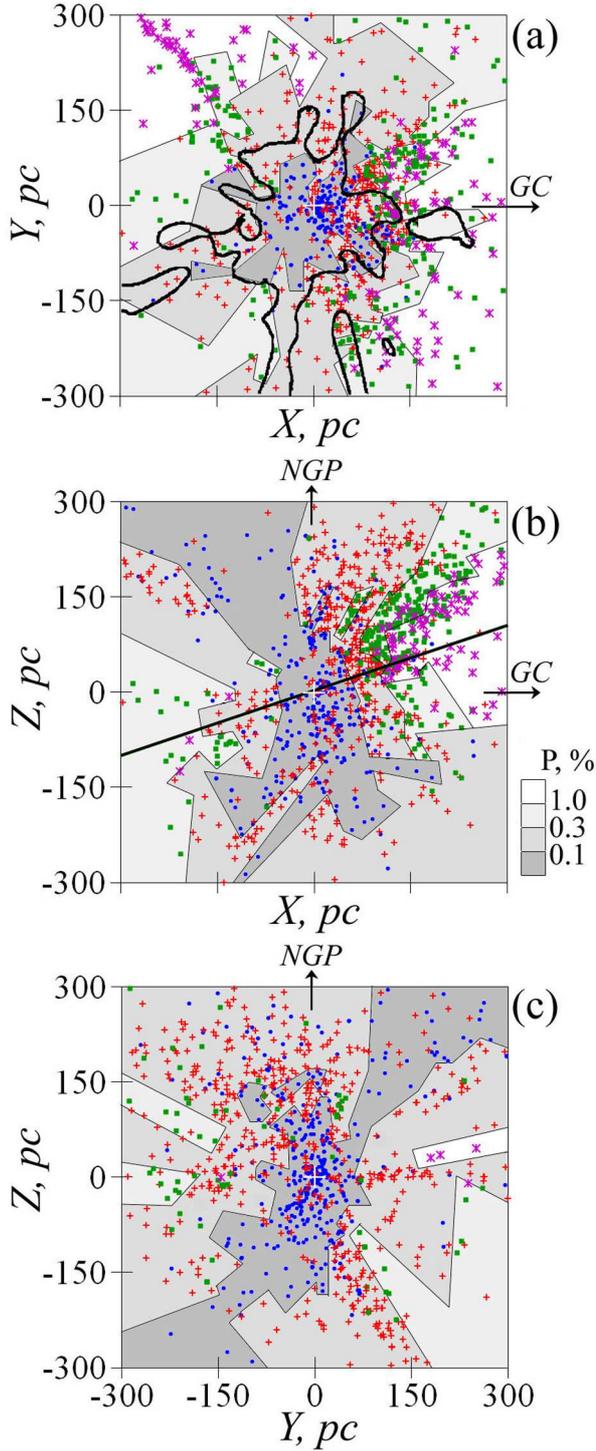}
\caption{The distribution of the sample along the $XYZ$ coordinates in the slices
(a) $-60<Z<60$ pc, (b) $-60<Y<60$ pc, and (c) $-60<X<60$ pc.
The stars with $P<0.1$, $0.1<P<0.33$, $0.33<P<1$ and $P>1$ per cent are shown by the 
blue dots, red crosses, green squares and purple snowflakes, respectively.
The regions with these different levels of $P$ are shown by shades of grey.
The black thick curve on the $XY$ plot is the \ion{Na}{i} volume density of $\log n_{\ion{Na}{i}}=-9.1$ cm$^{-3}$ 
iso-contour from \citet{welsh2010}.
The black thick line on the $XZ$ plot shows the orientation of the mid-plane of the Gould Belt.
The directions to the Galactic centre (GC) and the North Galactic pole (NGP) are shown.
}
\label{lowhigh}
\end{figure}

%¹¹¹¹¹¹¹¹¹¹¹¹¹¹¹¹¹¹¹¹¹¹¹¹¹¹¹¹¹
\begin{table}
\def\baselinestretch{1}\normalsize\small
\caption[]{The median $P$ (per cent) as a function of the Galactic octant from 1 to 8, latitude (degrees) 
and distance range (parsecs).
}
\label{pmap}
\[
\begin{tabular}{ccrrrr}
\hline
\noalign{\smallskip}
Octant & Latitude & \multicolumn{4}{c}{$P$, per cent} \\
\hline
\noalign{\smallskip}
 & & \multicolumn{4}{c}{Distance range, parsecs} \\
 & & 0--85 & 85--200 & 200--300 & 300--500 \\ 
\hline
\noalign{\smallskip}
1 & $b>+45$    & 0.06 & 0.19 & 0.25 & 0.24 \\
1 & $0<b<+45$  & 0.09 & 0.51 & 0.49 & 0.58 \\
1 & $-45<b<0$  & 0.09 & 0.31 & 0.44 & 0.73 \\ 	
1 & $b<-45$    & 0.01 & 0.08 & 0.12 & 0.08 \\
2 & $b>+45$    & 0.00 & 0.13 & 0.11 & 0.11 \\
2 & $0<b<+45$  & 0.00 & 0.17 & 0.18 & 0.19 \\ 	
2 & $-45<b<0$  & 0.09 & 0.18 & 0.23 & 0.31 \\ 	
2 & $b<-45$    & 0.09 & 0.14 & 0.17 & 0.18 \\ 	
3 & $b>+45$    & 0.06 & 0.08 & 0.11 & 0.11 \\
3 & $0<b<+45$  & 0.03 & 0.18 & 0.25 & 0.50 \\
3 & $-45<b<0$  & 0.05 & 0.15 & 1.07 & 1.22 \\ 	
3 & $b<-45$    & 0.02 & 0.11 & 0.23 & 0.19 \\
4 & $b>+45$    & 0.00 & 0.09 & 0.10 & 0.10 \\
4 & $0<b<+45$  & 0.22 & 0.21 & 0.14 & 0.18 \\ 	
4 & $-45<b<0$  & 0.08 & 0.30 & 0.56 & 0.71 \\ 
4 & $b<-45$    & 0.02 & 0.08 & 0.14 & 0.18 \\ 	
5 & $b>+45$    & 0.04 & 0.08 & 0.08 & 0.11 \\
5 & $0<b<+45$  & 0.05 & 0.10 & 0.08 & 0.11 \\
5 & $-45<b<0$  & 0.09 & 0.19 & 0.22 & 0.30 \\ 	
5 & $b<-45$    & 0.03 & 0.10 & 0.16 & 0.15 \\
6 & $b>+45$    & 0.04 & 0.13 & 0.12 & 0.11 \\
6 & $0<b<+45$  & 0.08 & 0.13 & 0.14 & 0.19 \\ 	
6 & $-45<b<0$  & 0.03 & 0.03 & 0.14 & 0.28 \\ 	
6 & $b<-45$    & 0.04 & 0.05 & 0.08 & 0.11 \\	
7 & $b>+45$    & 0.00 & 0.15 & 0.15 & 0.17 \\
7 & $0<b<+45$  & 0.01 & 0.15 & 0.25 & 0.29 \\ 
7 & $-45<b<0$  & 0.05 & 0.16 & 0.40 & 0.51 \\ 	
7 & $b<-45$    & 0.03 & 0.04 & 0.10 & 0.08 \\
8 & $b>+45$    & 0.04 & 0.19 & 0.24 & 0.31 \\ 	
8 & $0<b<+45$  & 0.09 & 0.30 & 0.52 & 0.54 \\ 	
8 & $-45<b<0$  & 0.07 & 0.16 & 0.35 & 0.55 \\ 	
8 & $b<-45$    & 0.02 & 0.10 & 0.17 & 0.23 \\ 	
\hline
\end{tabular}
\]
\end{table}

%¹¹¹¹¹¹¹¹¹¹¹¹¹¹¹¹¹¹¹¹¹¹¹¹¹¹¹¹¹¹¹¹¹¹¹¹

The final sample contains 4067 values of $P$ for 3871 stars and 4038 values of $\theta$ for 3842 stars.

The Hertzsprung--Russell (HR) diagram for our final sample is presented in Fig.~\ref{hr}, where
$G$, $G_{BP}$ and $G_{RP}$ are the {\it Gaia} DR2 photometric bands.
For 78 stars with no or uncertain $G$,
$G_{BP}$, and $G_{RP}$ they are calculated based on $B_T$ and $V_T$ photometry
from {\it Tycho-2} \citep{tycho2} by use of their relation from \citet{gaiaevans}.
We applied the reddening and extinction from \citet[][hereafter G17]{g17} taking into account the 
standard extinction law of \citet{ccm89} and spatial variations of $R_\mathrm{V}$ from its 3D map of \citet{rv}.
67 stars with high $P/E(B-V)$ are shown with the red diamonds.
They are discussed in Section~\ref{serkow}.
It is seen from Fig.~\ref{hr} that the giants and 
early classes of the main sequence are well-represented by our final sample.

The distribution of the sample in the sky by Galactic coordinates is shown in Fig.~\ref{lb}.
This distribution is quite representative, yet non-uniform.
The same is true for the 3D distribution of the sample in the Galactic $XYZ$ coordinates in Fig.~\ref{xyz}.
The stars with $P<0.1$ and $>0.1$ per cent are shown with the brown and blue symbols.
They draw the low and high polarisation in- and outside the Bubble, respectively.

\begin{figure}
\includegraphics{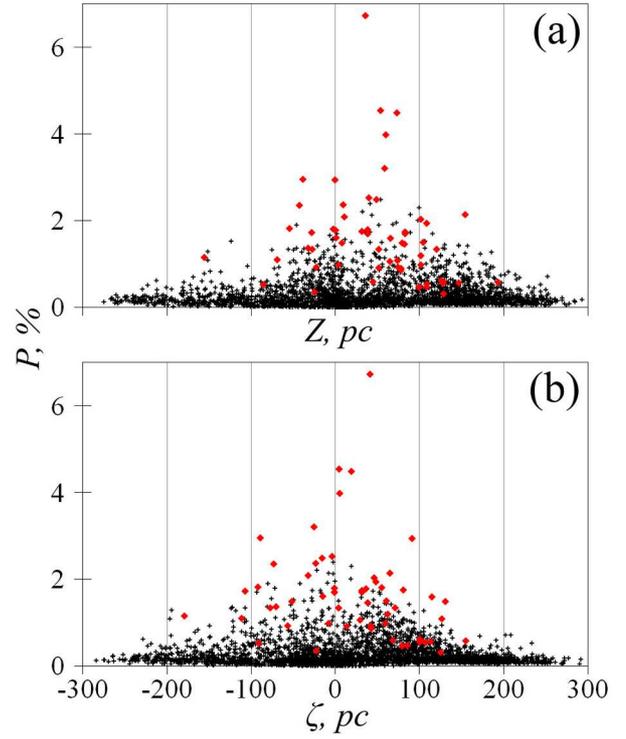}
\caption{The distribution of the subsample limited by $R<300$ pc along (a) $Z$ and (b) $\zeta$.
The 67 stars with $P/E(B-V)>9.3$ per cent per mag from all the reddening data sources are shown with the red diamonds.
The error bars are not shown because they are mostly smaller than the depicted symbols.
}
\label{zzeta}
\end{figure}

\subsection{Spatial variations of polarisation degree}
\label{degree}

The data under consideration are not enough to create a detailed 3D map of spatial variations of $P$ and $\theta$.
Yet, in Table~\ref{pmap} we provide a result of our study: a low resolution map where
the median of $P$ is given as a function of the Galactic octant, latitude, and distance range,
i.e. for 32 Galactic sectors and 4 distance ranges of 0--85, 85--200, 200--300, and 300--500 pc.

Due to the natural variations of the medium in every spatial cell and the lack of data, the accuracy of this map is 
limited.
This map would help to get a preliminary, first estimate of interstellar $P$ for a star. 

The value $P=0.1$ per cent can be accepted as a boundary between the regions with low and high polarisation
(the same value was accepted by \citet{leroy1999}).
Typically, the former are within 85 pc, i.e. nearly inside the Bubble, while the latter are outside.
Few exceptions are:
(i) in the $4^{th}$ octant, $0\degr<b<+45\degr$, several stars within $64<R<85$ pc have a high $P>0.1$ per cent and, hence, 
constitute a rather nearby wall of the Bubble;
(ii) several high latitude `chimneys' with low $P<0.1$ per cent are evident;
(iii) in the $5^{th}$ octant, $0\degr<b<+45\degr$ and $6^{th}$ octant, $-45\degr<b<0\degr$, there are two low latitude
elongations of the Bubble up to 200 or 300 pc with low $P<0.1$ per cent
(interpreted by \citealt{leroy1999} as a possible `chimney' to another bubble).

Table~\ref{pmap} shows that the variations of $P$ with $R$ are more noticeable than the ones with $l$ and $b$,
which are discussed in Section~\ref{colorbl}.
Fig.~\ref{pol} shows the variations of (a) $P$ and (b) $P/R$ with $R$ for all $l$ and $b$.
Despite a large scatter of these data, Fig.~\ref{pol} confirms our conclusions from Table~\ref{pmap}:
after a sharp increase of $P$ within nearly 100 pc, i.e. within the Bubble and its walls, 
outside it $P$ increases with $R$ gradually.
Fig.~\ref{pol}~(b) shows a drop of the average $P/R$ in- w.r.t. outside
the Bubble from 1.5 to 0.25 per cent per kpc, i.e. by about 6 times.
The maximum of $P/R$ at $R\approx150$ pc and its minimum at $R\approx450$ pc may suggest a different balance of 
polarisation and depolarisation of stellar radiation in these three distance ranges: $R<150$, $150<R<450$, and $R>450$ pc.
We note that 57 of 67 stars with $P/E(B-V)>9.3$ per cent per mag from all the reddening data sources,
shown with the open diamonds in Fig.~\ref{pol}, are within $98<R<314$ pc, i.e. not far from the maximum of $P/R$.
These stars are discussed in Section~\ref{serkow}.

In Fig.~\ref{lowhigh} we show three slices of the space under consideration:
(a) $X$ versus $Y$ with $-60<Z<60$ pc, (b) $X$ versus $Z$ with $-60<Y<60$ pc, and (c) $Y$ versus $Z$ with $-60<X<60$ pc.
The stars with $P<0.1$, $0.1<P<0.33$, $0.33<P<1$, and $P>1$ per cent are shown by different symbols.
The regions with these different levels of $P$ are shown by shades of grey.
The darkest tone approximately shows the Local Bubble in agreement with its other descriptions,
for example with the \ion{Na}{i} volume density of $\log n_{\ion{Na}{i}}=-9.1$ cm$^{-3}$ iso-contour from \citet{welsh2010}
shown in Fig.~\ref{lowhigh}~(a) as the black curve on the $XY$ plot.
In agreement with Table~\ref{pmap}, Fig.~\ref{lowhigh} shows the Bubble as a nearly spherical central cavity
with several low latitude elongations and
at least two wide `chimneys' to the Galactic halo in the northern 2$^{nd}$ and southern 4$^{th}$ quadrants.

Fig.~\ref{pol} and \ref{lowhigh}, together with Table~\ref{pmap}, 
may solve some contradictions in the description of this space.
For example, the note `the interstellar contribution [to polarisation] is not likely to be much more 
than 150 ppm [i.e. 0.015 per cent] even at 100 pc' \citep{cotton2016} seems to be controversial.
Now we see that $P=0.1$ per cent is a typical value at 100 pc.

The black line in Fig.~\ref{lowhigh}~(b) shows the orientation of the mid-plane of the Gould Belt.
It is tilted to the Galactic mid-plane by an angle of about 19$\degr$.
The Belt is rotated so that
its part with the highest $b\approx+19\degr$ is at $l\approx15\degr$, whereas its part with the lowest 
$b\approx-19\degr$ is at $l\approx195\degr$ \citep{gould, av}.
It is evident from both Fig.~\ref{lowhigh} and Table~\ref{pmap} that polarisation is higher along the Gould Belt,
i.e. in the 1$^{st}$ and 8$^{th}$ octants above the Galactic mid-plane and in the 4$^{th}$ and 5$^{th}$ octants below 
the plane.

Fig.~\ref{zzeta} shows the distribution of the subsample limited by $R<300$ pc along the coordinates 
(a) $Z$ and (b) $\zeta$, 
the distance from the mid-plane of the Gould Belt in the direction perpendicular to this plane.
$\zeta$ is an analogue of $Z$ in the reference frame of the Belt.
The definition of $\zeta$ is given by \citet{gould}:
\begin{equation}
\label{zeta}
\zeta=R(\cos\gamma\sin~b-\sin\gamma\cos~b\cos~l)\,,
\end{equation}
where $\gamma\approx19\degr$ is the tilt of the Belt to the Galactic mid-plane.
Fig.~\ref{zzeta} confirms that a maximum $P$ coincides with the mid-plane of the Gould Belt ($\zeta=0$ pc), 
but not with the Galactic mid-plane ($Z=0$ pc).
The shift of the distribution in Fig.~\ref{zzeta}~(a) to positive $Z$ is explained by a combination
of the growth of $P$ to both the mid-plane of the Belt and to the Galactic Centre.

We note a good agreement of our results with those from \citet{andersson2018} (see their Table~1), 
who have used the data from \citet{berd2014} and found a sharp increase of $P$ 
(`a wall of the Bubble') at rather high $b>+30\degr$ between the Galactic longitudes 350$\degr$ and 30$\degr$, 
where Fig.~\ref{lowhigh}~(b) shows plenty of green squares and purple snowflakes.
Apparently, this increase is due to the influence of the Gould Belt, which rises up to b=+19 deg and dominates 
within +10<b<+40 deg at these longitudes.
Yet, we note that our sample contains too little stars in this northern part of the Belt, i.e. near the upper part of the
black line in the $XZ$ plot of Fig.~\ref{lowhigh}.
This may lead to an underestimation of the importance of the Belt as a dust container and a 
domain of the spatial variations of $P$. More polarisation observations are needed here.

\begin{landscape}
\begin{figure}
\includegraphics{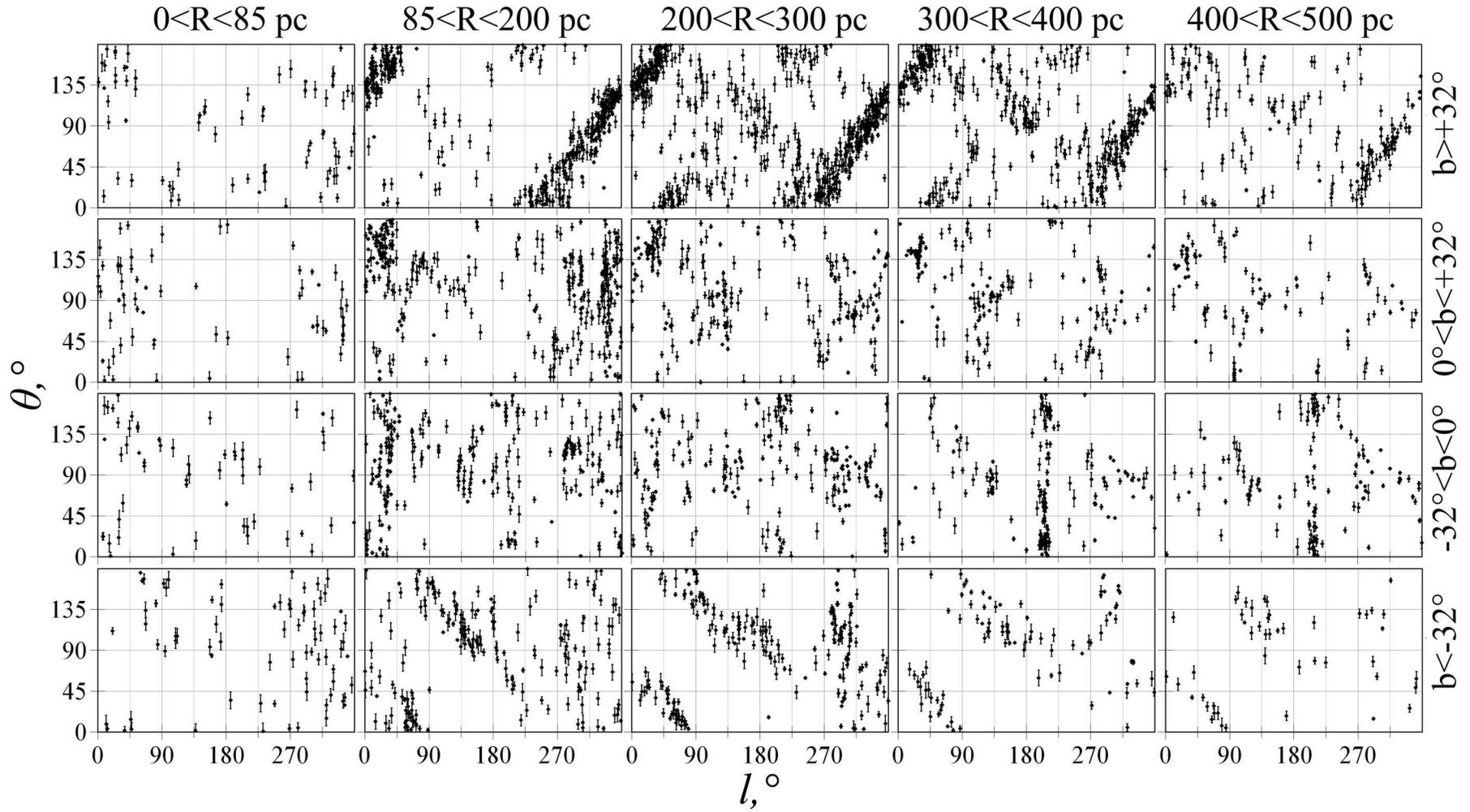}
\caption{$\theta$ versus $l$ for the different latitude and distance segments.
}
\label{pa}
\end{figure}
\end{landscape}

\subsection{Spatial variations of position angle}
\label{angle}

\begin{figure*}
\includegraphics{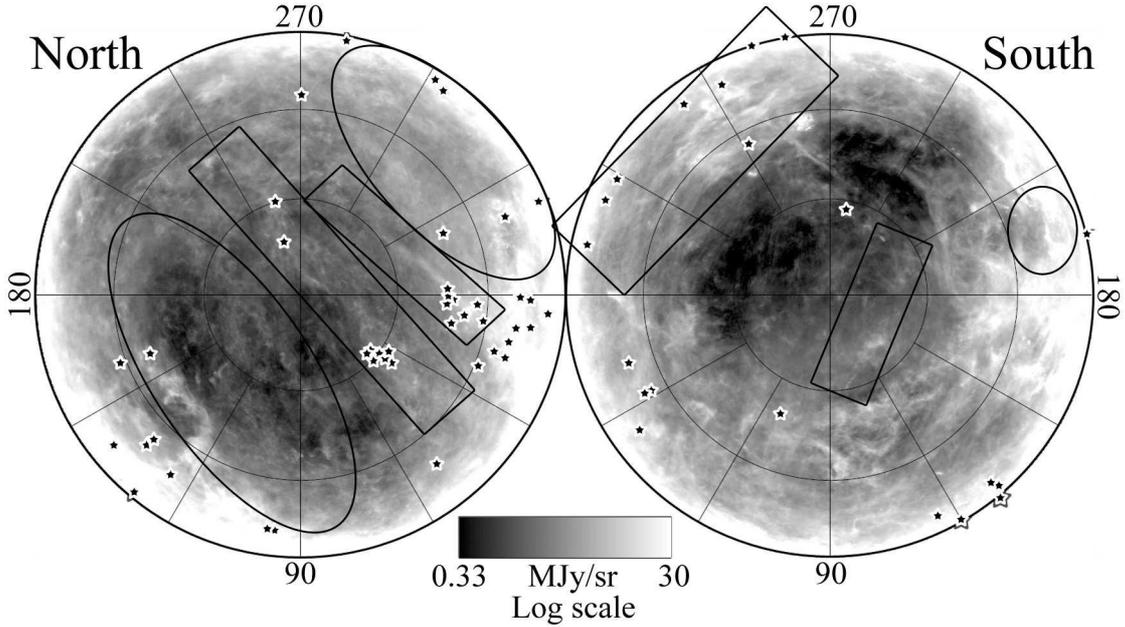}
\caption{The 67 stars with $P/E(B-V)>9.3$ per cent per mag from all the reddening data sources superimposed 
over the \citetalias{sfd} map.
Some high latitude areas of higher emission and reddening are shown by rectangles,
while some polarisation loops are shown by ovals.
}
\label{sfdfig}
\end{figure*}

We analysed the orientation of the polarisation vectors, i.e. their $\theta$, as a function 
of distance for 
four latitude zones: $b>+32\degr$, $0\degr<b<+32\degr$, $-32\degr<b<0\degr$, and $b<-32\degr$ in Fig.~\ref{pa}.
A similar analysis with comparable results has been carried out by \citet{stephens2011}.
The partitioning of the sky into these 4 zones is due to a rather different behaviour of $\theta$ in them.
Also, this partitioning is partially explained by a non-uniform distribution of the sample (see Fig.~\ref{lb}) and 
by the domination of the data from \citet{berd2014} at the high northern latitudes.

Almost no noticeable feature is seen inside the Local Bubble, i.e. in the leftmost plots of Fig.~\ref{pa}.
This cannot be explained by a larger uncertainty of $\theta$ due to a lower $P$ inside the Bubble:
the average $\sigma(\theta)$ is almost the same -- $6\degr$ and $5\degr$ for $R<85$ and $R>85$ pc, respectively.
Such various $\theta$ for one $l$ suggest a chaotic magnetic field and/or chaotic orientation of dust grains
and/or depolarisation of the polarised light by the radiation field of the Bubble.
The latter reason would be in line with the drop of $P/R$ within the Bubble seen in Fig.~\ref{pol}~(b).

In contrast, the space outside the Bubble is full of noticeable features.
In the northern 
high latitudes (the top plots) we see the same pattern for all $R$:
a regular bulk of stars draws a relation between $\theta$ and $l$ giving $\theta=0$ or $180\degr$ at $l\approx40\degr$ and 
$l\approx240\degr$
(the last value drifts from 240$\degr$ to 270$\degr$ with increasing $R$, as discussed below).
This relation points to a regular orientation of $\theta$ along a great loop with a radius of 
$(360-240+40)/2=80\degr$, i.e. nearly along a great circle.
This loop is located near the Nothern Galactic pole at $l\approx320\degr$ where $\theta\approx90\degr$.

This feature has been known since the first all-sky interstellar polarisation maps \citep{mathewson1970, leroy1999}.
The highest part of this loop at $b\approx+80\degr$ has long been known as the Markkanen's cloud of aligned dust 
\citep{markkanen1979}.
\citet{berd2002} have shown that Markkanen's cloud extends to lower latitudes, apparently, being a part of a large 
elongated dust envelope. 
The {\it Gaia} DR2 parallaxes are quite accurate in order to recognise 
the large thickness of this envelope: Fig.~\ref{pa} shows it at least for $85<R<400$ pc.
This agrees with a conclusion of \citet{berd2001} that the Markkanen's cloud starts at the distance of about 100 pc and 
extends well beyond 300 pc.
Moreover, this confirms the distance to the North Galactic Pole Rift, a well-defined \ion{H}{i} feature at about 100 pc 
\citep{puspitarini2012, snowden2015}, as a near part of the Markkanen's cloud.

\citet{berd2014} have shown that the Markkanen's cloud exactly follows a large bright filament 
(hereafter the Markkanen's filament)
of a higher dust far-IR emission from the emission/reddening map of \citet[][hereafter SFD]{sfd} 
(see figure~8 in \citeauthor{berd2014}).
The Markkanen's filament is delineated by the largest black rectangle in the centre of the left plot of 
Fig.~\ref{sfdfig} reproducing the \citetalias{sfd} map.

The next rectangle contains another filament in the 4$^{th}$ quadrant in Fig.~\ref{sfdfig}.
This filament coincides with the well-known North Polar Spur (NPS), one of the best known features in 
the radio continuum and diffuse soft X-ray background maps between $l\approx25\degr$, $b\approx+20\degr$ and
$l\approx330\degr$, $b\approx+75\degr$ \citep{lallement2016}.
This NPS filament does not enter the 3$^{rd}$ quadrant and has $\theta\approx0$ or 180$\degr$ at $l\approx270\degr$.
These properties allow us to find it as another region with aligned dust in Fig.~\ref{pa}.
As mentioned above, the intersection of the bulk of stars with $\theta=0$ or 180$\degr$ in the top plots drifts 
from $l\approx240\degr$ to 270$\degr$ when $R$ increases from 85 to 500 pc.
This means that the dominance of the Markkanen's filament in this bulk is changed by the dominance of the NPS filament
when $R>300$ pc.
This agrees with a conclusion of \citet{lallement2016}, who have been ruled out the Spur's near side closer 
than 300 pc.

Another regular bulk of stars is seen in Fig.~\ref{pa} at the high northern latitudes (the top plots) for
$50\degr<l<225\degr$ and $200<R<500$ pc.
It has an X-like shape in Fig.~\ref{pa}. This presents a loop of aligned dust delineated by the largest oval 
in the left plot of Fig.~\ref{sfdfig}.
The lowest latitude part of this loop is seen in the low northern latitude plots of Fig.~\ref{pa} as 
a minor bulk of stars at $90\degr<l<135\degr$, $200<R<400$ pc, and $\theta\approx90\degr$.

The last feature of aligned dust in the northern hemisphere is seen in Fig.~\ref{pa} 
as an X-like bulk of stars at the low northern latitudes, the 4$^{th}$ quadrant, $85<R<300$ pc.
This depicts another loop seen in Fig.\ref{sfdfig} as a filament delineated by an oval in the 4$^{th}$ quadrant.
As seen from the \ion{Na}{i} contour (the black
curve) in Fig.~\ref{lowhigh}~(a), this loop may be an interface between the Local and Loop I bubbles \citep{welsh2010}.
The Loop I Bubble is `80$\degr$ in radius and centred 120 pc away at $l=320\degr$, $b=5\degr$ for the neutral gas' 
\citep{frisch2007}.
\citet{leroy1999, reis2011, santos2011, frisch2015, cotton2017} discussed all the structures near this region
as an interface between the Local and Loop I bubbles.
In such a discussion one should take into account that Fig.~\ref{lowhigh} shows 
an elongation of the Local Bubble as a region with low polarisation both in this interface and in the place of 
the Loop I Bubble. However, Fig.~\ref{pa} shows that
the Local and Loop I bubbles are much smaller than their envelope of aligned dust.

This great envelope of the northern hemisphere contains the Markkanen's and NPS filaments and follows their
orientation.
Hence, this envelope is elongated between $l\approx40\degr$ and $250\degr$ and, thus, nearly follows a wide extension
of the Local Bubble at low latitudes of the 2$^{nd}$ and 6$^{th}$ octants, evident in Fig.~\ref{lowhigh}~(a) and 
known as the Local interstellar tunnel, or the Great tunnel \citep{welsh1991, av}.
It consists of the Cygnus Cavities, Local Bubble, and CMa tunnel \citepalias[see][figure 1]{lallement2014}.
This tunnel has many young stars and cloud complexes on its sides. Consequently, in fact, it defines the 
orientation of the Local spiral arm and its polarisation vectors.

As seen from Fig.~\ref{pa} and \ref{sfdfig}, the great envelope enters the low latitudes where it fades in a 
strong emission/reddening/extinction.
Yet, this envelope may re-appear as brightest filaments of the southern hemisphere.
However, Fig.~\ref{lb} shows that we have no enough data coverage of the largest southern filament, 
a curved one between the 1$^{st}$, 2$^{nd}$ and 3$^{rd}$ quadrants in the right plot of Fig.~\ref{sfdfig}.
Yet, two regular bulks of stars in Fig.~\ref{pa} are the filaments of aligned dust inside the rectangles in the right
plot of Fig.~\ref{sfdfig}.
The filaments in the
larger rectangle in the 4$^{th}$ quadrant are represented in Fig.~\ref{pa} by the bulk of stars 
with $\theta\approx90\degr$
at the low southern latitudes, $85<R<500$ pc.
In fact, Fig.~\ref{pa} shows similar regular bulks in the $2^d$ and $4^{th}$ quadrants at the low
northern and southern latitudes.
This suggests a prolongation of the discussed northern loops (the two ovals in the left plot of Fig.~\ref{sfdfig}) 
in the South.

The smaller rectangle in the 1$^{st}$, 2$^{nd}$ and 3$^{rd}$ quadrants in Fig.~\ref{sfdfig}
is represented in Fig.~\ref{pa} by a pronounced bulk of stars at the high southern latitudes, $85<R<400$ pc.
By their orientation and $\theta$ these southern filaments may be a prolongation of the great northern envelope.
More data are needed to make more robust conclusions on this envelope.

The last feature in the South is a compact loop in Orion delineated by the oval in the right plot of 
Fig.~\ref{sfdfig}.
It is seen in Fig.~\ref{pa} as a sharp vertical bulk of stars at $b\approx208\degr$ in the low southern latitudes
within $300<R<500$ pc. It has a chaotic $\theta$.

We note 67 stars with $P/E(B-V)>9.3$ per cent per mag shown in Fig.~\ref{sfdfig} by the star symbols.
Most of them appear at (i.e. before, in or behind)
the middle and high latitude filaments. We discuss these stars in Section~\ref{serkow}.

\section{Polarisation versus reddening}
\label{versus}

%¹¹¹¹¹¹¹¹¹¹¹¹¹¹¹¹¹¹¹¹¹¹¹¹¹¹¹¹¹
\begin{table}
\def\baselinestretch{1}\normalsize\small
\caption[]{The linear correlation coefficients of $P$ (per cent)
and $E(B-V)$ (mag) from the different reddening data sources and from {\it Gaia} DR2 
inside ($P<0.1$ per cent, the top right half of the Table) and 
outside ($P>0.1$ per cent, the bottom left half of the Table) the Bubble.
}
\label{correlation}
\[
\begin{tabular}{crrrrrrr}
\hline
\noalign{\smallskip}
 & \multicolumn{7}{c}{Inside} \\
\hline
\noalign{\smallskip}
 Outside & $P$ & \citetalias{g17} & \citetalias{av} & \citetalias{arenou} & \citetalias{drimmel} & \citetalias{lallement2014} & DR2 \\
\hline
\noalign{\smallskip}
$P$                        &      & 0.31 & 0.38 & 0.24 & 0.27    & 0.35 & $-0.06$ \\
\citetalias{g17}           & 0.59 &      & 0.64 & 0.48 & 0.46    & 0.59 & $-0.06$ \\
\citetalias{av}            & 0.50 & 0.76 &      & 0.67 & 0.57    & 0.71 & $-0.01$ \\
\citetalias{arenou}        & 0.56 & 0.73 & 0.69 &      & 0.38    & 0.46 & $0.00$ \\
\citetalias{drimmel}       & 0.08 & 0.16 & 0.20 & 0.04 &         & 0.49 & $0.01$ \\
\citetalias{lallement2014} & 0.66 & 0.80 & 0.69 & 0.72 & 0.02    &      & $0.04$ \\
DR2                        & 0.06 & 0.04 & 0.03 & 0.05 & $0.00$ & 0.07 &      \\
\hline
\end{tabular}
\]
\end{table}

%¹¹¹¹¹¹¹¹¹¹¹¹¹¹¹¹¹¹¹¹¹¹¹¹¹¹¹¹¹¹¹¹¹¹¹¹

%¹¹¹¹¹¹¹¹¹¹¹¹¹¹¹¹¹¹¹¹¹¹¹¹¹¹¹¹¹
\begin{table*}
\def\baselinestretch{1}\normalsize\small
\caption[]{The 67 stars with $P/E(B-V)>9.3$ per cent per mag by the reddening estimates from all the data sources:
the {\it Gaia} DR2 source\_id or {\it Hipparcos} number, $G$ magnitude, distance (parsecs), Galactic coordinates (degrees), 
$P$ and its $\sigma$ (per cents), $\theta$ and its $\sigma$ (degrees),
$R_\mathrm{V}$, and $E(B-V)$ from \citetalias{g17}, \citetalias{av}, \citetalias{arenou}, \citetalias{drimmel}, and
\citetalias{lallement2014}. The complete table is available online.
}
\label{higheff}
\[
\begin{tabular}{rlrrrrrrrrrrrrr}
\hline
\noalign{\smallskip}
DR2 Source\_id or HIP & $G$ & $R$ & $l$ & $b$ & $P$ & $\sigma(P)$ & $\theta$ & $\sigma(\theta)$ & $R_\mathrm{V}$ & G17 & G12 & AGG & DCL & LVV  \\
\hline
\noalign{\smallskip}
1037320780504503040 & 7.95 & 429 & 159.6896 & +37.8749 & 0.975 & 0.054 &  95 & 2 & 3.22 & 0.097 & 0.075 & 0.013 & 0.092 & 0.027 \\
1264203385691704448 & 8.96 & 150 &  36.7309 & +58.1949 & 0.560 & 0.028 & 172 & 1 & 3.59 & 0.040 & 0.056 & 0.015 & 0.028 & 0.032 \\
1264661680178370432 & 5.42 & 149 &  37.0262 & +59.2407 & 0.598 & 0.120 & 156 & 6 & 3.59 & 0.037 & 0.055 & 0.015 & 0.035 & 0.029 \\
1264797229346052864 & 4.15 & 114 &  36.2511 & +60.6605 & 0.478 & 0.029 & 150 & 2 & 3.59 & 0.005 & 0.050 & 0.027 & 0.026 & 0.020 \\
1265048261594583552 & 8.34 & 127 &  37.8240 & +58.8987 & 0.484 & 0.010 & 161 & 1 & 3.59 & 0.015 & 0.052 & 0.015 & 0.029 & 0.026 \\
\ldots & \ldots & \ldots & \ldots & \ldots & \ldots & \ldots & \ldots & \ldots & \ldots & \ldots & \ldots & \ldots & \ldots & \ldots \\
\hline
\end{tabular}
\]
\end{table*}

%¹¹¹¹¹¹¹¹¹¹¹¹¹¹¹¹¹¹¹¹¹¹¹¹¹¹¹¹¹¹¹¹¹¹¹¹

\subsection{Reddening data}
\label{reddata}

Reddening data sources, in contrast to polarisation data sources, are inhomogeneous in their origin and methods.
Their large systematic errors do not allow one to use them in many cases.
For example, only five of the thirteen selected publications on polarisation give a comparison between
the polarisation degree and the reddening.
\citet{heiles2000} considered $E(B-V)$, but with very low precision of 0.1~mag in almost all cases.
\citet{santos2011} found $E(b-y)$, but only for 5 per cent of their stars with polarisation measurements.
\citet{alves2006}, \citet{poidevin2006}, and \citet{andersson2007} used reddening and extinction estimates, 
but only in some small areas of the sky.

Recently, reddening estimates were obtained for millions lines of sight for the whole 
sky by use of far-IR dust emission and stellar multi-colour photometry. 
\citet{gm2017, gm2017big, gm2018}
have tested the random and systematic accuracy of various reddening data sources through
their ability to fit the precise photometry and parallaxes of 
the stars from the {\it Gaia} DR1 Tycho--Gaia Astrometric Solution \citep{tgas}
to some theoretical estimates in the HR diagram.
The theoretical estimates are based on the 
PARSEC \citep{bressan} and MIST \citep{mist} theoretical isochrones, and the TRILEGAL model of the Galaxy \citep{trilegal}.
\citet{gm2017, gm2017big, gm2018} concluded that the following 3D data sources (maps and models) are the most 
reliable within several hundred parsecs from the Sun:
\citet[][hereafter AGG]{arenou}, \citet[][hereafter DCL]{drimmel},
\footnote{The \citetalias{drimmel} estimates are calculated by use of the code of 
\citet{bovy2016}, \url{https://github.com/jobovy/mwdust}}
\citet[][hereafter G12]{gould, av},
and \citetalias{g17}.

In this study we use these data sources together with a new 3D map by \citetalias{lallement2014}, 
which may be
especially useful for the space under consideration.
\citetalias{lallement2014} provides the reddening only within $|Z|<300$ pc. 
Yet, since the Galactic dust layer lies mostly within $|Z|<300$ pc, 
we use the reddening at $Z=\pm300$ pc for $Z>\pm300$ pc along the same line of sight.

With the median $|Z|=122$ pc,
 most stars of our sample are located inside the inhomogeneous Galactic dust layer.
The authors of the selected reddening data sources have made some efforts to take it
into account \citep{gm2017big}. However, their median $E(B-V)$ for our sample are very diverse: 
0.087, 0.075, 0.048, 0.038, and 0.027~mag for
\citetalias{g17}, \citetalias{av}, \citetalias{arenou}, \citetalias{drimmel}, and \citetalias{lallement2014}, 
respectively.
The difference between these values is higher than the formal precision of these data sources.
Therefore, the following mutual analysis of $P$, $E(B-V)$, and $P/E(B-V)$ may
help to select the most reliable reddening estimates.

Reddening estimates from many other data sources are not applicable to stars inside the dust layer
within 500 pc from the Sun, as pointed out by \citet{gm2017big, gm2018}.
An example is the 2D (to infinity) reddening maps by \citetalias{sfd} and \citet{2015ApJ...798...88M},
based on the observations of the dust emission in far-IR by {\it COBE}, {\it IRAS}, and {\it Planck}.
The reduction of their reddening from infinity to a point inside the dust layer gives a reliable result only by 
use of either a model of the local dust spatial distribution (as realised in \citetalias{drimmel}), 
or some photometry of local stars.
Note that 3D reddening maps based on photometry of distant stars are not reliable in the nearby space.
An example is the map of \citet{green2018}, which provides a reliable reddening only for a minority of our stars.

{\it Gaia} DR2 also provides the reddening and extinction, but only for 68 per cent of our stars. 
Moreover, these reddenings and their very large scatter for close lines of sight seem to be unreliable.
For example, given the extinction law of \citet{ccm89}, {\it Gaia} DR2 gives for 658 stars at $|b|>50\degr$ the 
median $E(B-V)=0.131$~mag with its standard deviation $0.128$~mag
due to a natural diversity of the medium.
These values strongly disagree with the 
values $0.047 / 0.017$, $0.041 / 0.013$, $0.022 / 0.012$, $0.019 / 0.009$, and $0.012 / 0.008$~mag from
\citetalias{g17}, \citetalias{av}, \citetalias{arenou}, \citetalias{drimmel}, and \citetalias{lallement2014},
respectively, for the same stars.
Therefore, we do not use the Gaia DR2 reddening estimates.

\begin{figure}
\includegraphics{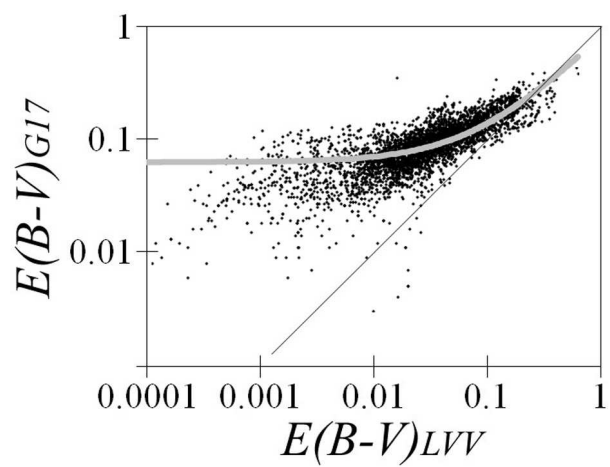}
\caption{$E(B-V)$ from \citetalias{lallement2014} versus \citetalias{g17} for our sample.
The linear trend is shown by the grey thick line, while the one-to-one relation -- by the black thin line.
}
\label{corr}
\end{figure}

However, we consider the reddening estimates from the {\it Gaia} DR2, recalculated into $E(B-V)$ by use of the 
extinction law of \citet{ccm89} with $R_\mathrm{V}=3.1$, among other estimates of $E(B-V)$ and the estimates of 
$P$, to calculate their linear correlation coefficients. They are presented in Table~\ref{correlation}.
The reddenings from the {\it Gaia} DR2 show
a very weak correlation with both $P$ and the remaining reddening estimates both in- and outside the Bubble.
The correlations between $P$, \citetalias{g17}, \citetalias{av}, \citetalias{arenou}, and \citetalias{lallement2014} 
are fairly tight, especially, outside the Bubble. Given
a rather significant differences between 
\citetalias{g17}, \citetalias{av}, \citetalias{arenou}, and \citetalias{lallement2014}, 
their high correlation coefficients mean
that their reddening estimates
can be basically described as some
linear trends of each other.
Indeed, Fig.~\ref{corr} shows $E(B-V)$ estimates from \citetalias{lallement2014} versus \citetalias{g17}.
The linear trend 
\begin{equation}
\label{g17lvv}
E(B-V)_{G17}=0.06+0.76\,E(B-V)_{LVV}\,,
\end{equation}
shown by the thick grey curve, reveals that 
\citetalias{lallement2014} and \citetalias{g17} differ mostly by the zero-point of 0.06~mag, most significant 
at high latitudes.
Besides this, Fig.~\ref{corr} shows a noticeable bulk of stars with $E(B-V)_{G17}<0.04$ and $E(B-V)_{LVV}<0.001$.
These are stars within 40 pc.
Their $E(B-V)_{LVV}$ seem to be well below the level which can actually be measured by photometry.
Yet, a low accuracy of both $P$ and $E(B-V)$ does not allow us to make any conclusion on this nearby space.

\begin{figure}
\includegraphics{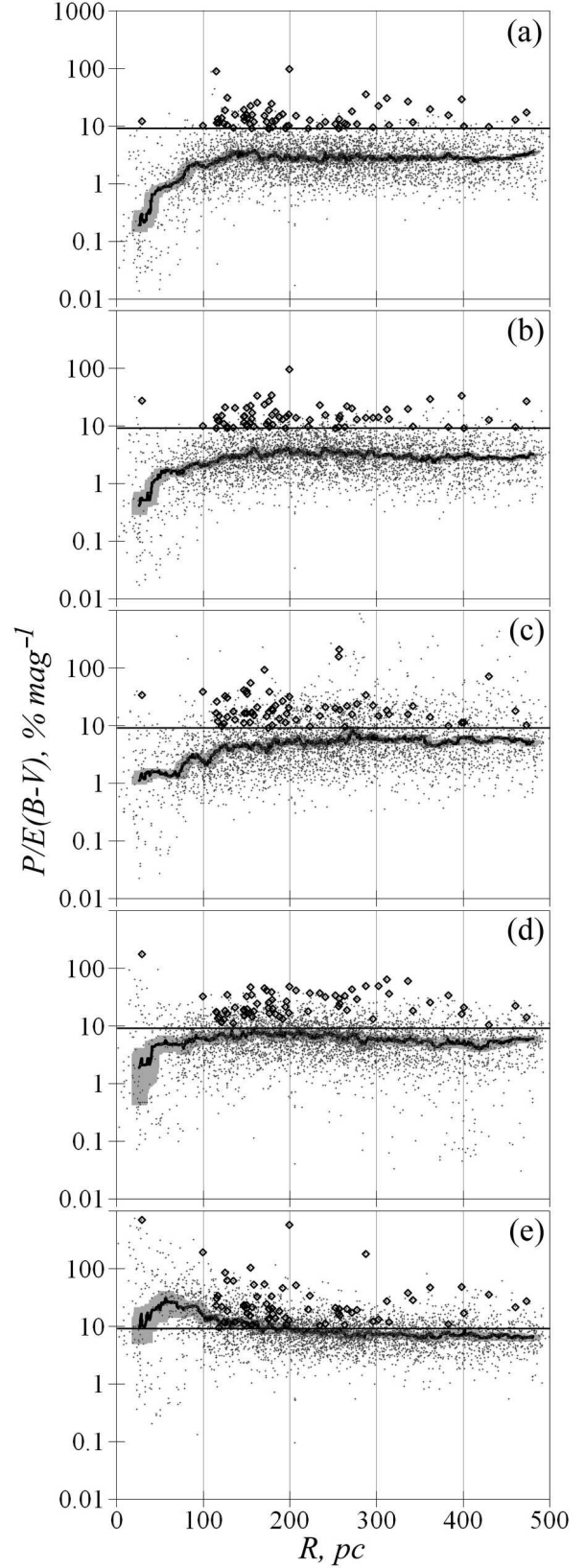}
\caption{$P/E(B-V)$ (per cent per mag) versus $R$ (parsecs) for the
(a) \citetalias{g17}, (b) \citetalias{av}, (c) \citetalias{arenou}, (d) \citetalias{drimmel}, 
and (e) \citetalias{lallement2014} $E(B-V)$ estimates.
All the curves are the moving medians over 101 points.
The 67 stars with $P/E(B-V)>9.3$ per cent per mag from all the reddening data sources are shown with the open diamonds.
}
\label{reff}
\end{figure}

\subsection{Dependences on $R$}
\label{onr}

Fig.~\ref{reff} shows $P/E(B-V)$ versus $R$ for 
(a) \citetalias{g17}, (b) \citetalias{av}, (c) \citetalias{arenou}, (d) \citetalias{drimmel}, 
and (e) \citetalias{lallement2014}.
It is seen that there is a slightly 
higher scatter of $P/E(B-V)$ for all the reddening data sources within nearly $R<100$ pc due to 
lower values and, consequently, 
higher fractional errors of both $P$ and $E(B-V)$
(see a discussion in \citet{frisch2015} regarding their figure 2).
Yet, Fig.~\ref{reff} shows a clear drop of $P/E(B-V)$ within nearly $R<100$ pc
by use of the data from \citetalias{g17}, \citetalias{av}, and \citetalias{arenou}.
Such a drop is not so evident for \citetalias{drimmel}, while \citetalias{lallement2014} clearly shows a monotonic
decrease of $P/E(B-V)$ with $R$ and a little depression of $P/E(B-V)$ only within $R<50$ pc,
in contrast to the other reddening data sources.
Comparing the spatial variations of $P$ and $P/R$ from Fig.~\ref{pol}, on the one hand, 
with the ones of $P/E(B-V)$ from Fig.~\ref{reff}, on the other hand,
we can conclude that \citetalias{lallement2014} shows a drop of $E(B-V)/R$ in- w.r.t. outside the Bubble 
similar to the one of $P/R$, 
whereas the remaining reddening data sources show no considerable drop of $E(B-V)/R$ at all.

Thus, choosing between the reddening estimates from \citetalias{lallement2014} or other data sources, 
we encounter a key problem for understanding the nature of the local interstellar medium,
whether $E(B-V)$ follows $P$ in its spatial variations.
We note that $E(B-V)$ and $P$ have different nature.
The latter is a percentage of the aligned dust, it is limited between 0 and 100 per cent, and 
it can either increase or decrease with $R$.
In contrast, $E(B-V)$ has no upper limit and cannot decrease with $R$.
Therefore, $E(B-V)$ and $P$ do not necessarily follow each other.

For 543 stars with $P=0.10\pm0.02$ per cent, i.e. at the boundary of the Bubble the median $E(B-V)$ are 
0.067, 0.056, 0.029, 0.025, and 0.014~mag from
\citetalias{g17}, \citetalias{av}, \citetalias{arenou}, \citetalias{drimmel}, and \citetalias{lallement2014}, 
respectively.
The diversity of these estimates is similar to the diversity of
the Bubble's boundary estimates mentioned in Section~\ref{intro}.
Therefore, given the nearly constant value of $P$ at the boundary of the Bubble,
we expect such a large diversity of the estimates of $P/E(B-V)$ to be with the median
1.5, 1.8, 3.4, 4.0, and 7.1 for 
\citetalias{g17}, \citetalias{av}, \citetalias{arenou}, \citetalias{drimmel}, and \citetalias{lallement2014}, 
respectively.
We discuss it in Section~\ref{serkow}.

Our results on the drops of $P/R$, $P/E(B-V)$ and, hence, $E(B-V)/R$ inside
the Bubble are compatible with the results of \citet{welsh2010}.
They have studied the equivalent width, as well as the column and volume density of \ion{Na}{i} and \ion{Ca}{ii} 
absorption lines towards 1857 stars.
By use of \ion{Na}{i} and \ion{Ca}{ii} they have determined the 3D spatial distribution of neutral and partly ionised
interstellar gas density, respectively.
Dust exists in both neutral and ionised regions.
But it is aligned to provide a considerable polarisation in the neutral regions only.
The Bubble is a space with many warm and partially ionised low density \ion{Ca}{ii}-bearing cloudlets
\citep{welsh2010}.
Hence, the \ion{Ca}{ii} marker should better correspond to the dust volume density and $E(B-V)/R$, 
whereas the \ion{Na}{i} marker -- to the polarisation degree.
This agrees with the conclusion of \citet{bailey2010} that `the dust responsible for the polarization is not located 
in the clouds responsible for the \ion{Ca}{ii} absorption.'
\citet{welsh2010} have revealed different drops of \ion{Na}{i} and \ion{Ca}{ii} volume densities inside
the Bubble (see their figures 9 and 10).
We have re-analysed their data by use of 
the {\it Gaia} DR2 parallaxes.
Taken $\varpi>14$ and $3<\varpi<7$~mas for the space in- and outside the Bubble, respectively,
we have found the drops by 6.2 and
1.5 times for \ion{Na}{i} and \ion{Ca}{ii} volume densities, respectively.
The former value agrees with the above mentioned drop of $P/R$,
whereas the latter value agrees with the negligible drop of $E(B-V)/R$ by use of all the reddening data sources,
except \citetalias{lallement2014}. Hence, the drop of $E(B-V)/R$ from \citetalias{lallement2014} by an order of 
magnitude makes them the outlier and leads to too many stars with too high $P/E(B-V)$.
Moreover, very high $P/E(B-V)$ in the Bubble from \citetalias{lallement2014} strongly contradicts to
the chaotic $\theta$ there, which is evident from Fig.~\ref{pa}.
Therefore, we are inclined to conclude that 
\citetalias{lallement2014} systematically underestimates $E(B-V)$ inside the Bubble.

It seems that the Bubble differs from the surrounding space by slightly lower volume density of dust 
(and $E(B-V)/R$ as its proxy) and about 6 times lower volume density of neutral gas (and $P/R$ as its proxy).
Yet, an overall gas-to-dust ratio in the Bubble can be maintained due to a several times higher volume density of 
ionised gas.
\citet{frisch2015} gave an example of a region inside the Bubble (BICEP2 field) where the stellar reddening suggests
the dust column density about 6.7 times higher than the neutral gas column density derived from the polarisation data.

Thus, a relation between the polarisation degree and reddening for any large volume of space with the
inhomogeneous medium inside, such as the equation~(\ref{fosa}) from \citet{fosalba2002}, has no sense.
Instead, one should consider such relations separately for every space region with homogeneous medium.

\subsection{Serkowski's limit}
\label{serkow}

\begin{figure}
\includegraphics{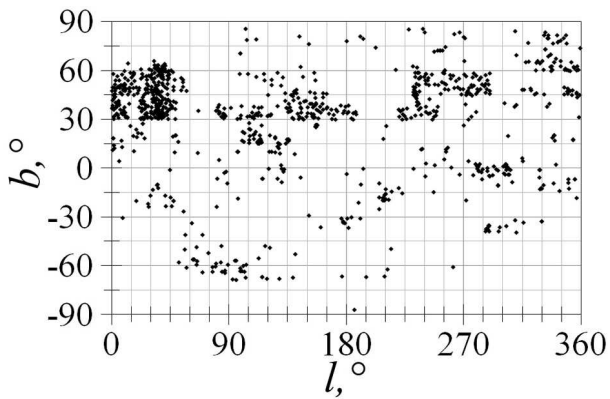}
\caption{The distribution of stars with $P/E(B-V)>9.3$ per cent per mag by use of \citetalias{arenou} in the sky.
}
\label{aggcells}
\end{figure}

The black horizontal lines in Fig.~\ref{reff} show the Serkowski's limit $P/E(B-V)=9.3$ per cent per mag.
It is seen that the reddening data sources give different number of cases with $P/E(B-V)>9.3$ per cent per mag: 
148 (3.8 per cent), 208 (5.4 per cent), 929 (24 per cent), 1033 (27 per cent), and 1660 stars (43 per cent)
for \citetalias{g17}, \citetalias{av}, \citetalias{arenou}, \citetalias{drimmel}, and 
\citetalias{lallement2014}, respectively.
Even among 3520 stars with $R>100$ pc, where the scatter of $P/E(B-V)$ is certainly lower, these numbers are
138 (3.9 per cent), 198 (5.6 per cent), 896 (25 per cent), 945 (27 per cent) and 1405 (40 per cent),
respectively.
These percentages can be compared with 4 per cent of such stars in \citet{planck21} and 
few per cents in the other studies.
By their very large percentage, 
\citetalias{arenou}, \citetalias{drimmel}, and \citetalias{lallement2014}, in fact, reject the Serkowski's limit.
Consequently, we must conclude that either Serkowski's limit is not valid at all, 
or \citetalias{arenou}, \citetalias{drimmel}, and \citetalias{lallement2014} 
considerably underestimate reddening within 500 pc, 
or $P/E(B-V)$ are strongly biased due to the selection effect in favour of stars with higher $P$.
The latter is quite unprobable, as follows from the agreed variations of $P$ in Fig~\ref{lowhigh}.

The stars with $P/E(B-V)>9.3$ per cent per mag have non-uniform distribution on the sky and along $R$.
For example, such stars selected by use of \citetalias{arenou} follow a coordinate partitioning during 
the creation of the \citetalias{arenou} model.
Namely, one can see in Fig.~\ref{aggcells} that the majority of these stars drop exactly to a few cells 
of this partitioning (see the Appendix of \citetalias{arenou}):
$15<b<30$ with $100<l<140$; 
$30<b<45$ with $0<l<50$;
$30<b<45$ with $80<l<190$;
$30<b<45$ with $220<l<250$;
$45<b<60$ with $0<l<60$;
$45<b<60$ with $110<l<170$;
$45<b<60$ with $230<l<290$; and
$b>60$ with $330<l<360$.
Most of these stars have $R>250$.
Apparently, the main reason of $P/E(B-V)>9.3$ per cent per mag in this case is an underestimation of $E(B-V)$ 
in these cells at rather large distances.
In turn, it is due to a poor sample of the distant stars used in the creation of \citetalias{arenou} in these cells.
\citetalias{drimmel} and \citetalias{lallement2014} show a higher percentage of the stars with 
$P/E(B-V)>9.3$ per cent per mag 
within $R<220$ pc and $R<150$ pc,
respectively, due to their very low reddening estimates near the Sun.
However, \citetalias{drimmel} admits this reddening as underestimated due to 
'the spatial resolution of the [DCL] dust model does not allow a detailed description of dust in the
vicinity of the Sun' \citepalias{drimmel},
while \citetalias{lallement2014} insists on such a very low reddening in uncompromising contrast to 
the Serkowski's limit.

Any case of $P/E(B-V)>9.3$ per cent per mag for a star by use of the only reddening data source can be 
explained by an underestimation of the reddening for this star by this data source.
Yet, we found 67 stars with $P/E(B-V)>9.3$ per cent per mag based on the reddening estimates from all five data sources. 
Our analysis of the SIMBAD data of these stars has revealed no indication of their intrinsic polarisation.
These stars are presented in Table~\ref{higheff} and shown 
in Fig.~\ref{hr}, \ref{lb}, \ref{pol}, \ref{zzeta}, \ref{sfdfig}, and \ref{reff} as the diamonds or stars.
It is seen from these figures that these stars represent all stellar classes and tend to lie in a few sky areas
and within the limited range $98<R<314$ pc (57 of 67 stars).
Fig.~\ref{sfdfig} shows these stars superimposed over the \citetalias{sfd} map.
It is seen that they tend to concentrate near the Galactic centre, the Markkanen's cloud, the North Polar Spur, 
and some filament dust clouds.
Probably, $P/E(B-V)>9.3$ per cent per mag for these stars is due to an underestimation of $E(B-V)$ in these regions.

On the other hand,
the correlation coefficient between $R_\mathrm{V}$ from \citet{rv} and $\log(P)$ for these stars is $-0.65$ 
(compared with a negligible correlation for the whole sample).
For most of these 67 stars with $P/E(B-V)>9.3$ per cent per mag we can obtain $P/A_\mathrm{V}<3$
(another version of the Serkowski's limit) by an acceptable deviation of $R_\mathrm{V}$ from the value 3.1.
Consequently, the too high $P/E(B-V)$ of these stars may be explained by a deviation of the extinction law 
near them from the standard one.

\citet{vos2016} have explained the existence of stars with $P/E(B-V)>9.3$ per cent per mag within their 
theoretical dust models considering oblate grains with axial ratio $>3$.
Such grains do not need to be perfectly aligned with the magnetic field to account for values above the Serkowski's limit.
\citeauthor{vos2016} even found $P/E(B-V)=9.5\pm2.5$ and $12.6\pm3.0$ per cent per mag for the Chamaeleon I and Musca 
clouds, respectively, by use of the data from \citet{andersson2007}.
However, \citeauthor{vos2016} found such an overcoming of the Serkowski's limit only for 2 out of 11 clouds within 500 pc, 
while \citetalias{lallement2014} suspect this to be for almost a half of the stars.
Therefore, this explanation seems to be valid only in some regions of the space.
As seen from Fig.~\ref{lb} and \ref{pol}, several our stars are in such a special region of Chamaeleon I and Musca clouds,
i.e. at $R=196$ pc, $l=297\degr$, $b=-15\degr$ and $R=171$ pc, $l=300\degr$, $b=-9\degr$, respectively.
Moreover, most or even all the 67 stars with $P/E(B-V)>9.3$ per cent per mag from all the reddening data sources
may be in such regions.
Anyway, the current data are not enough to exclude these stars from our study, especially, since
their influence on our results is negligible.

\subsection{Systematics of \citetalias{lallement2014}}
\label{syslvv}

Inexplicably large $P/E(B-V)$ from \citetalias{lallement2014} leads us to a discussion on their possible
systematic errors
(naturally, not excluding some systematic errors in the reddening data from the other sources).

The stellar sample used to create \citetalias{lallement2014} is far from being complete. For example, 
it contains less than 20,000 stars within $40<R<100$ pc,
as compared to a complete sample of about 50,000 turn-off stars used to create \citetalias{g17} in the same space.
Moreover, `There is a limitation in the brightness of the target stars, and the subsequent lack of strongly
reddened stars results\ldots There is for the same reasons a bias towards low opacities\ldots' 
\citepalias{lallement2014}.

\citetalias{lallement2014} determined the reddening as $E(b-y)=(b-y)-(b-y)_0$, where the intrinsic colour 
$(b-y)_0$ was calculated according to the spectral type of the star.
Intrinsic colours can be overestimated due to the well-known difficulties in the selection of 
unreddened stars.
\citetalias{lallement2014} stated the uncertainty $\sigma(E(B-V))\approx0.02$~mag for their data,
keeping in mind an offset of 0.02~mag found (and corrected) at the origin of the linear relationship between 
the Str\"omgren and Geneva photometric data.
However, such an additional {\itshape systematic} offset of $\Delta E(B-V)=0.02$~mag, if it is still hidden 
in the \citetalias{lallement2014} data, would explain a major part of its deviation from the rest data sources.
Such an offset would reconcile $E(B-V)_{G17}<0.04$ and $E(B-V)_{LVV}<0.001$ of the stars within 40 pc,
shown in Fig~\ref{corr} and discussed in Section~\ref{reddata}.

At high latitudes, behind the dust layer,
an offset of $\Delta E(B-V)=0.06$~mag would be more appropriate, as evident from Fig.~\ref{corr},
equation~(\ref{g17lvv}) and the discussion in Section~\ref{reddata}.
This would be in line with the conclusion of \citet{gm2018} that 
the median $0.04<E(B-V)<0.06$ through the dust half-layer at $|b|>50\degr$ gives the best agreement of the distribution
of the {\it Gaia} DR1 stars in the HR diagram with the theoretical predictions from PARSEC, MIST, and TRILEGAL,
mentioned in Section~\ref{reddata}.
\citetalias{lallement2014} noted 
`The solution is based not only on the color excess data but also on prior knowledge of the opacity 
distribution\ldots These two information sources complement each other: where the constraints from the data 
are insufficient, the inversion restores the prior density\ldots'.
At high latitudes, behind the dust layer,
the approach of \citetalias{lallement2014} sets the prior to $E(B-V)=0$,
then the sparse distribution 
of their high latitude stars (they even had to limit their map by $|Z|<300$ pc) 
keeps this zero prior for the final map.
This may be an explanation of a systematic underestimation of the reddening by \citetalias{lallement2014}.

\begin{figure}
\includegraphics{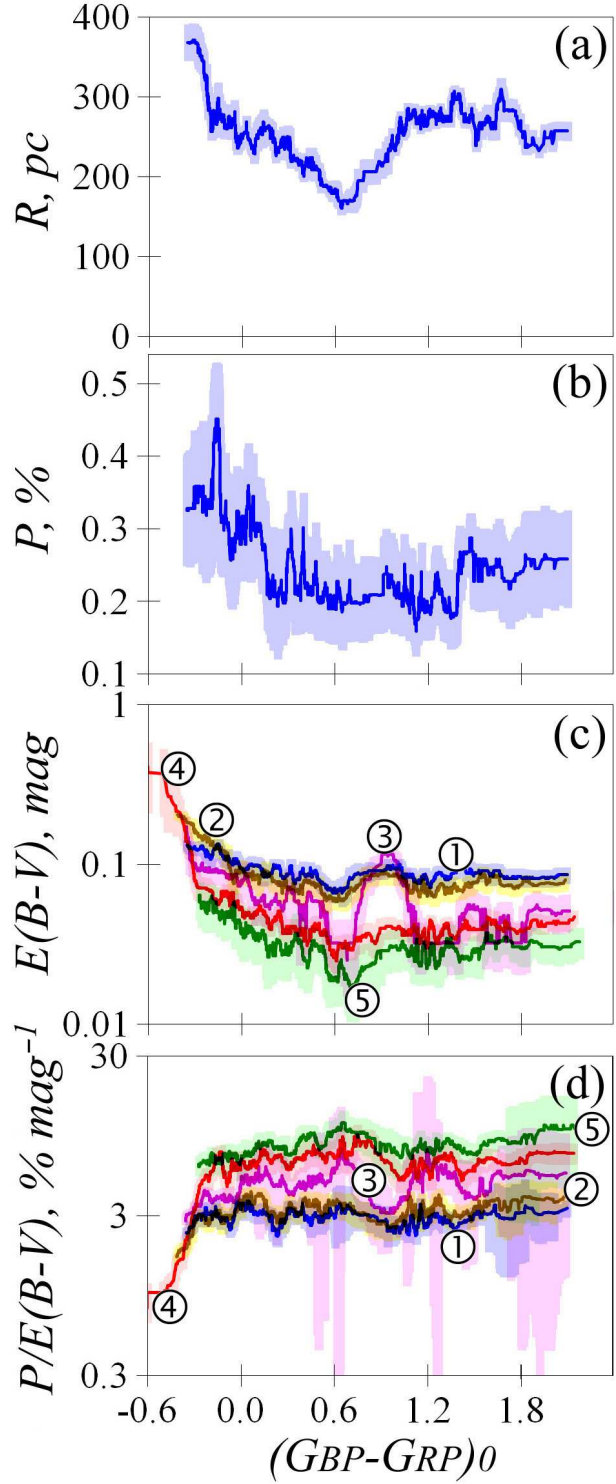}
\caption{(a) $R$ (parsecs), (b) $P$ (per cent), (c) $E(B-V)$ (mag) and (d) $P/E(B-V)$ (per cent per mag) 
versus the dereddened colour $(G_{BP}-G_{RP})_0$ (mag) for the stars with $R>100$~pc
as moving median over 101 points drawn by the blue, 
brown, purple, red and green curves with respecting $1\sigma$ variations as lighter hatched areas around them based on the
(1) \citetalias{g17}, (2) \citetalias{av}, (3) \citetalias{arenou}, (4) \citetalias{drimmel} and 
(5) \citetalias{lallement2014} data, respectively.
Only the results from \citetalias{g17} are shown in the plots (a) and (b) due to a similarity of the results from all 
the reddening data sources.
}
\label{color}
\end{figure}

\subsection{Dependences on colour and Galactic coordinates}
\label{colorbl}

In the following figures we show $P$, $E(B-V)$, and $P/E(B-V)$ versus the dereddened colour $(G_{BP}-G_{RP})_0$, 
$b$ and $l$ only outside the Bubble, since the results inside are less accurate.

Fig.~\ref{color} shows (a) $R$, (b) $P$, (c) $E(B-V)$ and (d) $P/E(B-V)$ versus the dereddened colour 
$(G_{BP}-G_{RP})_0$ for stars with $R>100$, based on the reddening estimates from 
\citetalias{g17}, \citetalias{av}, \citetalias{arenou}, \citetalias{drimmel}, and \citetalias{lallement2014}.

Fig.~\ref{color} shows some large variations of $R$, $P$, and $E(B-V)$ with colour.
However, these variations almost completely compensate each other, resulting in a nearly constant $P/E(B-V)$ as
a function of colour.
Particularly, the well-known higher $P$ of OB and M stars, seen in Fig.~\ref{color}~(b), 
is explained by their higher $R$ and compensated by their higher $E(B-V)$.
This is true for all the reddening data sources, despite very different estimates of $E(B-V)$ and $P/E(B-V)$
for the same colour by various data sources.
Moreover, \citetalias{drimmel} even shows a
significant decrease of $P/E(B-V)$ for OB stars w.r.t. other classes.
Yet, this is due to their very high $E(B-V)$, which can be erroneous.

\begin{figure}
\includegraphics{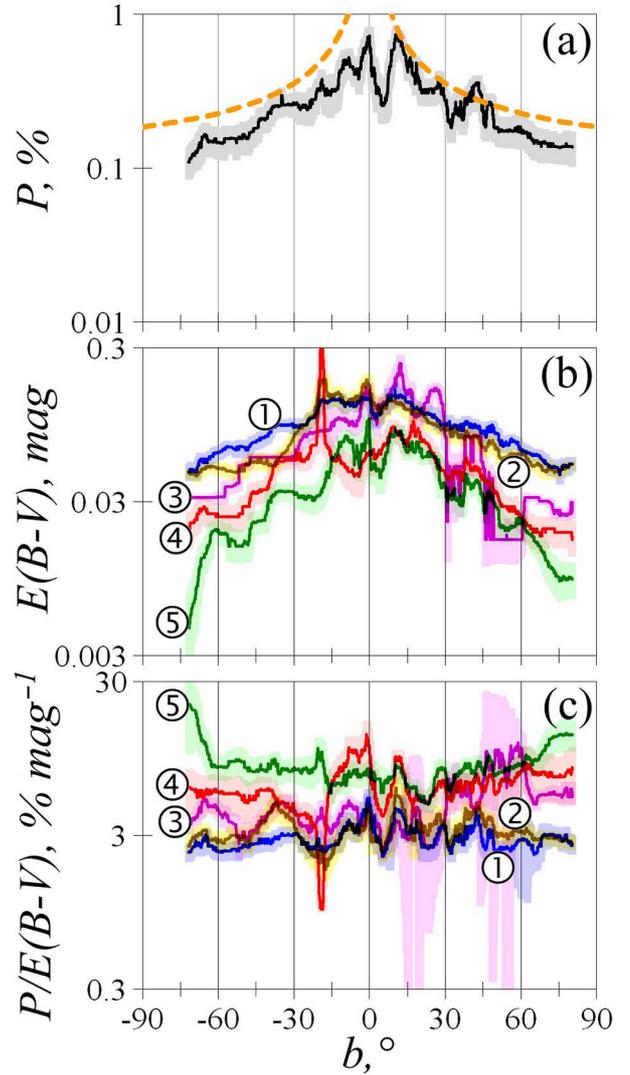}
\caption{(a) $P$ (per cent), (b) $E(B-V)$ (mag) and (c) $P/E(B-V)$ (per cent per mag) 
versus $b$ (degrees) for the stars with $R>100$~pc
as moving median over 101 points drawn by the blue, brown, purple, red and green curves with
respecting $1\sigma$ variations as lighter hatched areas around them based on 
(1) \citetalias{g17}, (2) \citetalias{av}, (3) \citetalias{arenou}, (4) \citetalias{drimmel} and 
(5) \citetalias{lallement2014} data, respectively.
The orange dashed curve is a function found by \citet{fosalba2002}.
}
\label{lat}
\end{figure}

\begin{figure}
\includegraphics{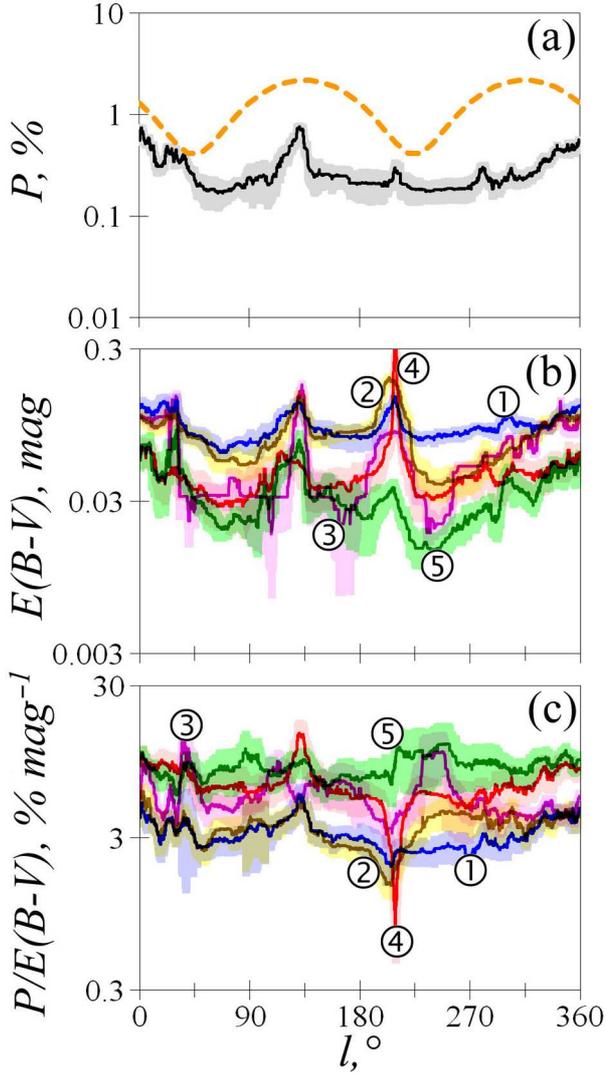}
\caption{The same as Fig.~\ref{lat} but for the stars with $P>0.1$ per cent at all the distances, and $l$ as the argument.
The orange dashed curve is a function found by \citet{fosalba2002}.
}
\label{lon}
\end{figure}

Fig.~\ref{lat} shows (a) $P$, (b) $E(B-V)$, and (c) $P/E(B-V)$ versus $b$ for the stars with $R>100$~pc, 
based on the reddening estimates from 
\citetalias{g17}, \citetalias{av}, \citetalias{arenou}, \citetalias{drimmel}, and \citetalias{lallement2014}.
Both $P$ and $E(B-V)$ show a rise within $|b|<20\degr$ due to the influence of the Gould Belt, as a dust container. 
There is also a sharp peak near $b\approx-19\degr$ due to a large number of stars near the directions to
the Orion clouds ($l\approx208\degr$), where the Belt has a large deviation from the Galactic equator.

The orange dashed curve shows an approximating function obtained by \citet{fosalba2002}.
A rather poor agreement on this curve with our data is due to 
the fact that \citet{fosalba2002} considered all the stars within several kiloparsecs from the Sun.

As for the dependence on colour in Fig.~\ref{color}~(d), it is seen from Fig.~\ref{lat}~(c) that
the large variations of $P$ and $E(B-V)$ with $b$ almost completely compensate each other,
resulting in a nearly constant $P/E(B-V)$, albeit at some different levels for different reddening data sources.
Some residual variations of $P/E(B-V)$ on $b$ can be explained:
for \citetalias{arenou} -- by very low accuracy at high latitudes, which is seen in the very wide error belt;
for \citetalias{drimmel} -- by a non-uniform spatial distribution of our sample.
\citetalias{lallement2014} is the only reddening data source showing a significant increase of $P/E(B-V)$ with $|b|$. 
Since it has no natural explanation, it should be considered as a confirmation
of the above mentioned systematic underestimation of $E(B-V)$ at high latitudes by \citetalias{lallement2014}.

For the dependences on colour and $b$, we use the limit $R>100$ pc.
We do not use the limit $P>0.1$ per cent instead, i.e. in order to select the space 
exactly outside the Bubble, because it excludes the high latitude `chimneys' with low $P$.
This would result in a poor statistics on $b$ and a bias in the statistics on some stellar classes 
due to their different distribution on $b$.
However, the dependences on $l$ appears to be more sensitive to the boundaries of the Bubble than the 
dependences on colour and $b$.
This is expected after a note of \citet{bailey2010}:
'The relationship between polarization and angle to the line of sight should lead to a polarization that depends on 
galactic longitude'.
Therefore, Fig.~\ref{lon} shows (a) $P$, (b) $E(B-V)$, and (c) $P/E(B-V)$ versus $l$ for the stars with $P>0.1$ per cent
at all $R$ for the same data as in Fig.~\ref{lat}.

The dependences in Fig.~\ref{lon}~(a) and (b) are defined by the increases of both $P$ and $E(B-V)$ from all the 
data sources 
(i) about the direction to the Galactic centre, as well as in the Gould Belt about its 
(ii) higher ($l\approx15\degr$, $b\approx19\degr$) and 
(iii) lower ($l\approx195\degr$, $b\approx-19\degr$) deviations from the Galactic mid-plane \citep[see][ Table~4]{av}.
This results in two wide smooth increases: 
a higher one at $l\approx0\degr$ and a lower one at $l\approx180\degr$.
This is
seen in all reliable local reddening maps and models, at least, beginning with Arenou et al. 1992 (see, 
for example, Figure~2 in \citet{gould}).
Moreover, a non-uniform distribution of our sample makes some regions of the Belt more pronounced. 
For example, a lot of stars of the 
Perseus-Cassiopea ($l\approx130\degr$, $b\approx0\degr$) and Orion ($l\approx208\degr$, $b\approx-19\degr$) complexes 
in the sample give the related spikes in Fig.~\ref{lon}.

A double sinusoid obtained by \citet{fosalba2002} is shown in Fig.~\ref{lon} by the orange dashed curve.
Its extrema do not fit our data due to the consideration by \citeauthor{fosalba2002} of many stars with $R>500$ pc, 
i.e. behind the Belt.

As Fig.~\ref{color} and \ref{lat}, Fig.~\ref{lon} shows that some large variations of $P$ and $E(B-V)$ with $l$ 
almost completely compensate each other,
resulting in a nearly constant $P/E(B-V)$, albeit at some different levels for different reddening data sources.

\section{Conclusions}
\label{conclusions}

We have analysed an all-sky compilation of the precise optical interstellar polarisation data from thirteen data 
sources for 
3871 {\it Gaia} DR2 and {\it Hipparcos} stars within 500 pc from the Sun and shown them versus five 3D maps and models of 
reddening $E(B-V)$, the most reliable ones within this space.
We excluded from consideration the stars with a known or suspected intrinsic polarisation.
The comparison of the polarisation data from different data sources for common stars has 
shown that these data can be compiled in one sample.

We have created a map, where the median $P$ is given for a grid of 8 Galactic octants per 2 hemispheres per 
2 latitude zones per 4 distance ranges within 500 pc.
The value $P=0.1$ per cent can be accepted as a value which separates the space inside the Local Bubble with
a lower $P$ and the space outside with a higher one.
$P$ is maximal at the mid-plane of the Gould Belt, but not at the Galactic mid-plane.

In the Bubble, $P/R$ shows the same drop by about 6 times, as the drop of the volume density of \ion{Na}{i}
and neutral gas.
In contrast, all the reddening data sources, except \citetalias{lallement2014}, show a negligible
drop of $E(B-V)/R$ and, hence, of dust volume density, which is similar to the drop by about 1.5 times
of the volume density of \ion{Ca}{ii}.
Thus, the Bubble seems to differ
from the surrounding space by slightly lower volume density of dust and several times lower volume density of neutral gas.
An overall gas-to-dust ratio in the Bubble can be maintained due to a several times higher volume density of 
ionised gas.
Finally, it seems that the Bubble is not a space of a lower volume density of matter, but the one of a higher 
ionisation of it.

In contrast to the other reddening data sources, \citetalias{lallement2014} shows in the Bubble 
a drop of $E(B-V)/R$ by an order of magnitude, i.e. comparable with the drop of $P/R$.
Consequently, \citetalias{lallement2014} gives $P/E(B-V)$ higher in- than outside the Bubble, 
while the remaining reddening data sources -- vice versa.
This result of \citetalias{lallement2014} strongly contradicts to the chaotic $\theta$ in the Bubble.

Our analysis shows that the addition of a nearly constant systematic offset would eliminate most discrepancies
between \citetalias{lallement2014} and the other reddening data sources.

The 3D reddening map of \citetalias{g17} and 3D extinction model by \citetalias{av}
give an acceptable percentage of stars overcoming the Serkowski's limit $P/E(B-V)<9.3$ per cent per mag.
The 3D map/models by \citetalias{arenou}, \citetalias{drimmel}, and \citetalias{lallement2014} give too many cases
with $P/E(B-V)>9.3$ per cent per mag, leading us to the conclusion that
either Serkowski's limit is not valid at all, or these maps/models considerably underestimate reddening 
within some regions.
These regions mostly have $R<220$, $<150$ and $>250$ pc for \citetalias{drimmel}, \citetalias{lallement2014}, and
\citetalias{arenou}, respectively.
For \citetalias{arenou} these regions are related to the partitioning of the sky during the creation of this model.

We found 67 stars with $P/E(B-V)>9.3$ per cent per mag based on the reddening estimates from all five data sources. 
These stars tend to concentrate near the Galactic centre, the Markkanen's cloud, the North Polar Spur, 
and some filament dust clouds.
Probably, $P/E(B-V)$ is high for these stars due to an unusual extinction law, 
an underestimation of $E(B-V)$ in these regions, or an influence of oblate dust grains.

Outside the Bubble, the spatial variations of $P$ and its variations with dereddened colour $(G_{BP}-G_{RP})_0$ are
almost compensated by the variations of $E(B-V)$, resulting in a nearly constant $P/E(B-V)$.
Hence, the well-known higher values of $P$ for OB and M stars are completely compensated by their higher $R$ and $E(B-V)$.

We found a chaotic behaviour of $\theta$ inside the Bubble.
In contrast, the middle and high northern latitudes outside the Bubble are dominated by a giant envelope of aligned dust
with a length of $>120\degr$, a width of $>50\degr$, and a depth of $>300$ pc.
This envelope is elongated between $l\approx40\degr$ and $250\degr$, i.e. nearly along the Local interstellar tunnel.
The well-known Markkanen's cloud of aligned dust and the North Polar Spur belong to this envelope, 
while the two giant loops of aligned dust in the second and fourth quadrants seem to edge the envelope.
All these structures are found as some emission/reddening filaments on the \citetalias{sfd} map. 
Probably, this giant northern envelope continues at the low and southern latitudes.

\section*{Acknowledgements}

We thank the reviewers Dr. Barry Welsh and Dr. Ralf Siebenmorgen for their useful comments.
We thank Dr. Andrei Berdyugin for a discussion of his valuable polarisation results.
We thank Dr. Ruslan Yudin and Dr. Vladimir Grinin for useful discussion of our results.
The resources of the Centre de Donn\'ees astronomiques de Strasbourg, Strasbourg, France 
(\url{http://cds.u-strasbg.fr}), including the SIMBAD database, were widely used in this study.
This work has made use of data from the European Space Agency (ESA) mission {\it Hipparcos}.
This work has made use of data from the European Space Agency (ESA) space mission {\it Gaia} 
(\url{https://www.cosmos.esa.int/gaia}), 
processed by the {\it Gaia} Data Processing and Analysis Consortium 
(DPAC, \url{https://www.cosmos.esa.int/web/gaia/dpac/consortium}).
This research has made use of the Washington Double Star Catalog maintained at the U.S. Naval Observatory.

\bsp	% typesetting comment
\label{lastpage}
\end{document}